\documentclass[letterpaper]{article}
\usepackage[totalwidth=480pt, totalheight=680pt]{geometry}

\usepackage[T1]{fontenc}
\usepackage{amsfonts,amsmath,amssymb,amsthm}
\usepackage{algorithm}
\usepackage{algorithmicx}
\usepackage{algpseudocode}
\usepackage{booktabs}
\usepackage{tabularx}
\usepackage{url}
\usepackage[round]{natbib}
\usepackage{xcolor}
\definecolor{darkgreen}{rgb}{0,0.5,0}
\usepackage{hyperref}
\hypersetup{
    colorlinks=true,
    linkcolor=red,
    citecolor=darkgreen,
    filecolor=magenta,
    urlcolor=blue
}
\usepackage{authblk}

\newtheorem{theorem}{Theorem}[section]
\newtheorem{proposition}[theorem]{Proposition}
\newtheorem{lemma}[theorem]{Lemma}
\newtheorem{corollary}[theorem]{Corollary}
\theoremstyle{definition}
\newtheorem{definition}[theorem]{Definition}
\theoremstyle{remark}
\newtheorem{remark}[theorem]{Remark}

\makeatletter
\providecommand*{\theHALG@line}{}
\renewcommand*{\theHALG@line}{\thealgorithm.\arabic{ALG@line}}
\makeatother

\DeclareMathOperator{\OPT}{OPT}
\DeclareMathOperator{\NSW}{NSW}
\newcommand{\F}{\mathcal{F}}
\newcommand{\cA}{\mathcal{A}}
\newcommand{\cO}{\mathcal{O}}

\title{\bf Computing Feasible EF1 Allocations under\\ Generalized Assignment Constraints}
\author[1]{Eugene Lim}
\author[2]{Nicholas Teh}

\affil[1]{National University of Singapore, Singapore}
\affil[2]{University of Oxford, UK}
\date{}

\begin{document}

\maketitle

\begin{abstract}
We study the fair allocation of indivisible goods under generalized assignment constraints, where each agent's bundle must fit within her budget, the sizes of goods may differ across agents, and unallocated goods go to a charity. A natural fairness notion here is feasible envy-freeness up to one good (FEF1), which requires an agent's envy towards any feasible subset of another bundle to disappear after some good is removed from this subset. While FEF1 allocations always exist, they could previously be computed efficiently only when sizes are common to all agents. We show that FEF1 allocations can be computed in strongly polynomial time, and more generally whenever each agent's value can be almost exactly maximized over her feasible bundles, a requirement that cannot be relaxed substantially. Our analysis relies on an essentially tight quadratic bound on how often each agent's bundle is repaired. Moreover, the allocation can be made approximately envy-free up to any good while retaining nearly half the Nash welfare of any given allocation. In contrast, requiring a single good to eliminate an agent's envy towards all feasible subsets of another bundle at once, which makes no difference without constraints, renders the problem NP-hard and may halve the attainable Nash welfare.
\end{abstract}

\section{Introduction}

The fair allocation of resources among interested agents is a fundamental problem in society, and has been studied in mathematics, economics, and more recently in computer science \citep{brams1996fair,moulin2003fair,amanatidis2023fair}. While the majority of the fair division literature assumes that any bundle of resources can be given to any agent, in many applications an agent can only accept a limited amount. A branch of a museum has limited floor space for the exhibits that it receives, and a reviewer has limited time for the papers assigned to her. Moreover, the amount of space or time that an item consumes often depends on its recipient---an exhibit may require a dedicated room in one branch but fit into an existing gallery in another, and an expert reads a paper in her area faster than a newcomer does. Such situations are captured by \emph{generalized assignment constraints}, named after the classical generalized assignment problem in combinatorial optimization \citep{shmoys1993gap}: each agent has her own size for every item and a budget on the total size of her bundle \citep{barman2023gap}. When the budgets are tight, it may be impossible to allocate every item; following the literature, we place the unallocated items in a \emph{charity} \citep{wu2021budget,chaudhury2021charity}.

In order to reason about fairness, we must define when an allocation is considered to be ``fair''. A prominent notion is \emph{envy-freeness}, which requires every agent to like her bundle at least as much as the bundle of any other agent. Under constraints, however, an agent may envy a bundle that she could not accept herself---a branch with little floor space has no use for a collection of exhibits that it cannot display. It is therefore natural to compare an agent's bundle only with those subsets of another bundle that she could accommodate. Accordingly, an allocation is said to be \emph{feasibly envy-free} if no agent prefers such a subset of another agent's bundle, or of the charity, to her own bundle \citep{wu2021budget,barman2023gap}. The comparison with the charity is essential: without it, leaving every item unallocated would be trivially fair. Unfortunately, feasible envy-freeness cannot always be satisfied when the items are indivisible. Indeed, if a single valuable item fits within the budgets of two agents, then no matter how the item is allocated, at least one of the agents will be left envying the bundle that contains it, regardless of whether this bundle belongs to the other agent or to the charity. A natural relaxation is \emph{feasible envy-freeness up to one good (FEF1)}, which requires that, for every subset of another bundle that an agent can accommodate, her envy towards this subset can be eliminated by removing some good from it \citep{wu2021budget,cohen2026budget}.

Under \emph{budget constraints}, where every item has the same size for all agents, the computation of FEF1 allocations has received considerable attention. \citet{wu2021budget} introduced this model and showed that maximizing the Nash welfare yields a $1/4$-approximation of FEF1. For identical valuations, \citet{gan2023approximation} computed $1/2$-approximate FEF1 allocations in polynomial time, as well as exact FEF1 allocations when all agents have the same budget or there are only two agents \citep[see also][]{wu2025budget}. Under the same assumption, \citet{barman2023budget} computed allocations in which the envy disappears after the removal of two goods---even when sizes differ across agents---as well as FEF1 allocations in several special cases, and identified generalized assignment constraints as an interesting direction for future work.

\citet{barman2023gap} then showed that, under generalized assignment constraints, an allocation satisfying the stronger notion of \emph{FEFx}---which requires the envy towards every feasible subset of another bundle to disappear after the removal of any good from this bundle---always exists, so FEF1 allocations exist as well. However, their algorithm runs in pseudo-polynomial time, and they proved that computing an FEFx allocation is NP-hard. The polynomial-time computation of FEF1 allocations, mentioned as an open question by \citet{barman2023budget} and \citet{elkind2024chores}, was recently achieved by \citet{hu2026ef1} for budget constraints. Their algorithm assumes that sizes are common to all agents, and its running time depends on the magnitudes of the numbers in the input. This raises the question of whether FEF1 allocations can also be computed efficiently under generalized assignment constraints, and if so, whether the running time can be made independent of the numbers in the input.

A subtlety in answering this question is that the term FEF1 has been used for two different notions. In the notion described above, the removed good may depend on the subset being compared. \citet{dror2021matroid} and \citet{amanatidis2025submodular} instead require a single good whose removal from the other bundle eliminates the envy towards all feasible subsets of the remainder at once; we refer to this notion as \emph{strong FEF1}. In the absence of constraints, the comparison with an entire bundle is the most demanding one, so both notions reduce to the well-known notion of envy-freeness up to one good (EF1) \citep{lipton2004approximately,budish2011combinatorial}. Already under budget constraints, however, the two notions differ. For instance, suppose that an agent can accommodate any two goods and receives a good of value $100$, while four goods that she values at $51$ each remain in the charity. Every pair of goods in the charity is worth $102$ to her, but removing either good of the pair eliminates her envy, so FEF1 holds; on the other hand, no matter which good is removed from the charity, a pair worth $102$ remains, so strong FEF1 fails. FEF1 treats each feasible subset of another bundle as a separate alternative to the agent's bundle, which fits the view that she could have taken only one such subset. Strong FEF1 instead asks for a single good that accounts for her envy towards all of these alternatives at once, as EF1 does for an entire bundle in the absence of constraints. As we shall see, this difference determines whether the notion can be achieved efficiently, and can even halve the attainable Nash welfare.

In this paper, we answer this question in the affirmative: we show that FEF1 allocations can be computed in strongly polynomial time under generalized assignment constraints, and more generally under any downward-closed constraints that admit sufficiently accurate optimization oracles; we also show that the required accuracy cannot be relaxed substantially. Moreover, FEF1 allocations can be computed so as to retain nearly half of the Nash welfare of any given allocation. By contrast, we prove that computing an allocation satisfying strong FEF1 is NP-hard, both when there is only one agent and when strong FEF1 is only required between agents.

\subsection{Our Results}

As is very common in fair division, we assume that the agents have nonnegative additive valuations, and we require fairness both between agents and towards the charity. We describe the model formally in Section~\ref{sec:prelim}, where we also show that FEFx implies strong FEF1, which in turn implies FEF1, and that neither implication can be reversed (Proposition~\ref{prop:hierarchy}). Our results fall into three groups: an efficient algorithm and its analysis (Section~\ref{sec:algorithm}), the limits of tractability (Section~\ref{sec:hardness}), and guarantees on the Nash welfare (Section~\ref{sec:welfare}).

In Section~\ref{sec:algorithm}, we consider FEF1. Our algorithm follows the framework of \citet{hu2026ef1}: starting with all goods in the charity, it repeatedly gives some agent a subset of the charity that she values considerably more than her current bundle, after shrinking this subset until no agent values any of its proper subsets that she can accommodate considerably more than her own bundle. Since the agents' utilities never decrease, the guarantees established when a bundle is created remain valid while the bundle stays with its recipient, and the algorithm stops once no agent values any feasible subset of the charity considerably more than her own bundle. Although these guarantees are only approximate, removing the most valuable good from a subset compensates for the approximation, which is why approximate optimization suffices. We replace their shrinking procedure by one that examines each agent only once, which reduces the number of oracle calls per update from $O(nm^2)$ to $O(n)$, where $n$ and $m$ denote the numbers of agents and goods. We also observe that the correctness argument never relates the feasibility of a set for one agent to its feasibility for another agent, so the algorithm returns an FEF1 allocation whenever every agent's feasible bundles form a downward-closed family over which her value can be approximately maximized.

Our main technical contribution is an analysis of the number of updates that does not depend on the numbers in the input. Besides the \emph{ordinary updates} described above, the framework of \citet{hu2026ef1} performs \emph{repairs}: if some agent who can accommodate the shrunken subset would violate FEF1 towards it, she receives this subset without a good of minimum value to her. \citet{hu2026ef1} showed that each agent receives at most $m^3$ repairs, and their bound on the number of ordinary updates grows with the logarithm of the ratio between an agent's total value and her smallest positive value. We show that each agent receives at most $m(m-1)/2$ repairs, and that, for a suitable approximate oracle, our algorithm may indeed perform $\Omega(m^2)$ repairs (Theorem~\ref{thm:repairs} and Proposition~\ref{prop:repair-lower}). The upper bound rests on the observation that every repair moves the agent's utility past a sum of values that are consecutive in the sorted order of her values. We also show that each agent receives $O(m^2)$ ordinary updates (Lemma~\ref{lem:ordinary}). Consequently, the algorithm makes $O(nm^2)$ updates and $O(n^2m^2)$ oracle calls (Theorem~\ref{thm:oracle}).

Under generalized assignment constraints, a profit-scaling scheme for the knapsack problem can be used as the oracle, so an FEF1 allocation can be computed in $O(n^2m^5)$ time---in particular, in strongly polynomial time (Corollary~\ref{cor:gap}). Even for budget constraints, this improves upon the running time of $O(n^2m^9+n^2m^7(L+\ln m))$ established by \citet{hu2026ef1}, where $L$ is the length of the input. By adjusting the precision of the oracle, the computed allocation can be made $(1-\eta)$-FEFx at the same time for any given $\eta>0$, which complements the approximation scheme of \citet{barman2023gap}, since a $(1-\eta)$-FEFx allocation need not be FEF1. Beyond generalized assignment constraints, our algorithm applies whenever each agent's constraints form the intersection of two matroids (Remark~\ref{rem:matroid}).

In Section~\ref{sec:hardness}, we show that these positive results are close to the limits of tractability. Although allocations satisfying strong FEF1 always exist, we show that computing one is NP-hard, even when there is only one agent and her value for each good equals its size (Theorem~\ref{thm:hardness}). We then show that the strong comparison with the charity is not the source of this hardness: computing an allocation that satisfies strong FEF1 between agents and FEF1 towards the charity is NP-hard as well (Theorem~\ref{thm:between}). The difficulty here is that the bundle in which the hard comparison must take place is chosen by the algorithm rather than by the reduction. We overcome this with a polynomial-size family of agents whose sizes are defined through affine maps over a finite field, so that every sufficiently large bundle contains, for at least one of these agents, two goods of each size used in the reduction (Lemma~\ref{lem:affine}). Both hardness results persist if any fixed number of goods may be removed instead of one (Corollary~\ref{cor:k-removals}), and they strengthen the NP-hardness of FEFx due to \citet{barman2023gap}. Finally, we show that the accuracy of the optimization required by our algorithm cannot be dispensed with. FEF1 itself becomes NP-hard when each agent's feasible bundles are the common independent sets of three partition matroids, even for one agent with unit values (Theorem~\ref{thm:matroid3}), in contrast to the tractability for two matroids. Moreover, a polynomial-time approximation scheme for the optimization problem does not suffice: computing an FEF1 allocation is NP-hard for one agent who has budgets for three different resources (Theorem~\ref{thm:budgets3}), although maximizing her value subject to these budgets admits such a scheme. In the oracle model of Section~\ref{sec:algorithm}, we show that an accuracy of $1-c/m$ for any constant $c>2$ requires exponentially many queries (Theorem~\ref{thm:queries}), so the accuracy of $1-1/(2m)$ required by our algorithm is optimal up to a constant factor in the error.

In Section~\ref{sec:welfare}, we turn to efficiency. Fairness alone does not guarantee that the goods are put to good use---for instance, if every agent can accommodate only one good, then leaving all goods in the charity is FEF1. We measure efficiency by the Nash welfare, i.e., the geometric mean of the agents' utilities, and show that any given allocation can be converted into one that is FEF1 and $(1-\eta)$-FEFx while retaining a $1/2-\varepsilon$ fraction of its Nash welfare, in time polynomial in $n$, $m$, $1/\varepsilon$, and $1/\eta$ (Theorem~\ref{thm:nsw} and Corollary~\ref{cor:nsw-gap}). By a result of \citet{feldman2024tradeoffs}, the factor $1/2$ cannot be improved as $\eta$ tends to zero, even in the absence of constraints. This upper bound relies on the $(1-\eta)$-FEFx guarantee; without constraints, for instance, an EF1 allocation of maximum Nash welfare always exists \citep{caragiannis2019unreasonable}. Letting $\varepsilon$ and $\eta$ tend to zero, we also obtain that an FEFx allocation retaining half of the Nash welfare of any given allocation always exists (Corollary~\ref{cor:fefx-exist}), which extends a result of \citet{caragiannis2019donating} to arbitrary downward-closed constraints. Combined with an approximation algorithm for the Nash welfare under submodular valuations, applied to a fractional relaxation of the agents' constraints, Theorem~\ref{thm:nsw} yields an allocation that is FEF1 and $(1-\eta)$-FEFx and whose Nash welfare is at least a $1/(12+\varepsilon)$ fraction of the optimum (Theorem~\ref{thm:nsw-opt}). Finally, we show that the order of quantifiers matters for welfare as well: even in instances where some FEF1 allocation maximizes the Nash welfare and all budgets are almost equal, every allocation satisfying strong FEF1 between agents may retain only about half of the maximum Nash welfare (Theorem~\ref{thm:nsw-strong}). By Corollary~\ref{cor:fefx-exist}, this is the largest possible loss.

A summary of our results can be found in Table~\ref{tab:results}.

\begin{table}[t]
\caption{Complexity of computing allocations with charity for nonnegative additive valuations, with $n$ agents and $m$ goods. The upper part concerns generalized assignment constraints, where allocations satisfying each notion always exist, since FEFx allocations do \citep[Theorem~2]{barman2023gap}. FEFx implies strong FEF1, and every notion other than $(1-\eta)$-FEFx implies FEF1 (Proposition~\ref{prop:hierarchy}). ``NSW'' refers to the Nash welfare.}
\label{tab:results}
\begin{center}
\small
\renewcommand{\arraystretch}{1.25}
\begin{tabularx}{\textwidth}{@{}p{.36\textwidth} X p{.2\textwidth}@{}}
\toprule
Notion & Computation & Reference \\
\midrule
FEFx & NP-hard; pseudo-polynomial time & \citet[Thms.~3,~5]{barman2023gap} \\
$(1-\eta)$-FEFx & Polynomial in the input length and $1/\eta$ & \citet[Thm.~4]{barman2023gap} \\
Strong FEF1 & NP-hard, even for one agent & Theorem~\ref{thm:hardness} \\
Strong FEF1 between agents, FEF1 towards the charity & NP-hard & Theorem~\ref{thm:between} \\
FEF1 and $(1-\eta)$-FEFx & $O(n^2(m^5+m^3/\eta^2))$ time & Corollary~\ref{cor:gap} \\
FEF1, $(1-\eta)$-FEFx, and a $1/2-\varepsilon$ fraction of the NSW of a given allocation & Polynomial in $n$, $m$, $1/\varepsilon$, and $1/\eta$ & Corollary~\ref{cor:nsw-gap} \\
FEF1, $(1-\eta)$-FEFx, and a $1/(12+\varepsilon)$ fraction of the optimal NSW & Polynomial in the input length, $1/\varepsilon$, and $1/\eta$ & Theorem~\ref{thm:nsw-opt} \\
FEF1 & Strongly polynomial, $O(n^2m^5)$ time & Corollary~\ref{cor:gap} \\
 & Polynomial time for budget constraints & \citet{hu2026ef1} \\
\midrule
FEF1, intersections of two matroids & Strongly polynomial time & Remark~\ref{rem:matroid} \\
FEF1, intersections of three partition matroids & NP-hard, even for one agent with unit values & Theorem~\ref{thm:matroid3} \\
FEF1, $3$-dimensional budget constraints & NP-hard, even for one agent & Theorem~\ref{thm:budgets3} \\
FEF1, $(1-c/m)$-approximate oracles, $c>2$ & Exponentially many queries, even for one agent & Theorem~\ref{thm:queries} \\
\bottomrule
\end{tabularx}
\end{center}
\end{table}

\subsection{Related Work}

Constrained fair division has attracted substantial interest in recent years; we refer to the surveys by \citet{suksompong2021constraints} and \citet{biswas2023structured} for an overview. Budget constraints were introduced by \citet{wu2021budget}, and subsequent work has studied approximation algorithms \citep{gan2023approximation,barman2023budget}, envy-freeness up to any good \citep{dai2022efx}, the price of fairness \citep{hu2026pof}, chores \citep{elkind2024chores}, and online arrivals \citep{cohen2026budget}. Besides computing FEF1 allocations, \citet{hu2026ef1} showed that deciding whether a given allocation is FEF1 is coNP-complete, even for one agent, gave polynomial-time algorithms for envy-freeness up to any good in two special cases, and showed that, for identical valuations, their algorithm can be made to retain a $1/2-\varepsilon$ fraction of the maximum utilitarian welfare. Their hardness concerns checking an allocation supplied as part of the input, whereas our results concern allocations chosen by the algorithm. For divisible goods, \citet{barman2023gap} showed that feasibly envy-free allocations exist under generalized assignment constraints and can be computed efficiently, and \citet{barman2026divisible} studied their compatibility with Pareto optimality and truthfulness.

Matroid constraints form another prominent class of downward-closed constraints. \citet{biswas2018cardinality} computed EF1 allocations under cardinality constraints, \citet{dror2021matroid} studied strong FEF1 for complete allocations under heterogeneous matroid constraints, and \citet{cookson2025constrained} examined the compatibility of EF1 with Pareto optimality under matroid constraints. For submodular valuations, \citet{amanatidis2025submodular} computed allocations satisfying approximate strong FEF1 under $p$-system constraints. \citet{li2021hereditary} studied maximin share guarantees for valuations induced by hereditary set systems. Conflict constraints, under which an agent may not receive two goods that are in conflict with each other, are also downward-closed \citep{hummel2022conflict}.

The Nash welfare has played a central role in fair division since \citet{caragiannis2019unreasonable} showed that maximizing it yields EF1 allocations for additive valuations. The maximum Nash welfare rule has since been characterized as the only welfarist rule that guarantees EF1 \citep{suksompong2023mnw,yuen2023mnw}, and this characterization persists even when EF1 is weakened to envy-freeness or proportionality up to any fixed number of goods \citep{celine2026mnw}. Leaving goods unallocated has proved useful for combining fairness with efficiency: \citet{caragiannis2019donating} showed that a partial allocation that is envy-free up to any good and retains half of the maximum Nash welfare always exists, \citet{chaudhury2021charity} used a small charity to obtain envy-freeness up to any good, and \citet{feldman2024tradeoffs} characterized the optimal tradeoff between approximate envy-freeness up to any good and the Nash welfare. When all goods must be allocated and the number of goods exceeds the number of agents by at most three, \citet{lim2026efx} showed that the loss of Nash welfare due to envy-freeness up to any good vanishes as the number of agents grows. \citet{barman2024subadditive} extended such guarantees to subadditive valuations, and our approach in Section~\ref{sec:welfare} adapts theirs. Under budget constraints, \citet{garbea2023efx} combined a feasible version of envy-freeness up to any good with a constant fraction of the maximum Nash welfare for two and three agents. For submodular valuations, \citet{garg2023nsw} approximated the maximum Nash welfare within a factor of $4+\varepsilon$, and \citet{bei2026nsw} improved this factor to $3.56+\varepsilon$ using a randomized algorithm. The minimal envied subsets of \citet{chaudhury2021charity}, which \citet{barman2023gap} adapted to generalized assignment constraints, also underlie the candidate construction in our algorithm.

\section{Preliminaries}
\label{sec:prelim}

Let $N=[n]$ be the set of agents and $M$ be the set of $m\ge 1$ indivisible goods, where we use $[z]$ to denote $\{1,2,\dots,z\}$ for each positive integer $z$. Each agent $i\in N$ has a nonnegative additive valuation, i.e., a value $v_i(g)\ge 0$ for each good $g\in M$, and $v_i(S)=\sum_{g\in S}v_i(g)$ for each $S\subseteq M$.

Each agent $i$ has a family $\F_i\subseteq 2^M$ of \emph{feasible} bundles, which we assume to be \emph{hereditary} (or downward-closed): $\varnothing\in\F_i$, and if $S\in\F_i$ and $T\subseteq S$, then $T\in\F_i$. Under \emph{generalized assignment constraints}, agent $i$ has a size $s_i(g)>0$ for each good $g$ and a budget $B_i\ge 0$, and a bundle $S$ is feasible for her if and only if its total size $s_i(S):=\sum_{g\in S}s_i(g)$ is at most $B_i$. \emph{Budget constraints} form the special case in which $s_i(g)=s_j(g)$ for all agents $i,j$ and goods $g$ \citep{wu2021budget}. For each agent $i$ and set $D\subseteq M$, let $\OPT_i(D):=\max\{v_i(S): S\subseteq D,\ S\in\F_i\}$ be the largest value that agent $i$ can feasibly obtain from $D$.

An \emph{allocation} $\cA=(A_1,\dots,A_n)$ is a collection of pairwise disjoint bundles such that $A_i\in\F_i$ for every agent $i$, and the \emph{charity} $C:=M\setminus\bigcup_{i\in N}A_i$ consists of the unallocated goods. We write $u_i:=v_i(A_i)$ for the utility of agent $i$. From the perspective of agent $i$, a \emph{target} is either the bundle $A_j$ of another agent $j\ne i$ or the charity $C$. Following \citet{wu2021budget} and \citet{barman2023gap}, we compare an agent's bundle only with the subsets of a target that are feasible for her.

\begin{definition}[FEF1]
\label{def:fef1}
An allocation is \emph{feasibly envy-free up to one good (FEF1)} if, for every agent $i$, every target $D$, and every nonempty $T\subseteq D$ with $T\in\F_i$, there exists a good $g\in T$ such that $u_i\ge v_i(T\setminus\{g\})$.
\end{definition}

Since valuations are additive, this condition is equivalent to $u_i\ge v_i(T)-\max_{g\in T}v_i(g)$ for every such set $T$. Under budget constraints, FEF1 is the notion studied, under the name EF1, by \citet{wu2021budget}, \citet{gan2023approximation}, and \citet{hu2026ef1}; \citet{cohen2026budget} used it under generalized assignment constraints.

\begin{definition}[Strong FEF1]
\label{def:strong}
An allocation satisfies \emph{strong FEF1} if, for every agent $i$ and every nonempty target $D$, there exists a good $g\in D$ such that $u_i\ge v_i(T)$ for every $T\subseteq D\setminus\{g\}$ with $T\in\F_i$.
\end{definition}

Equivalently, strong FEF1 requires that $u_i\ge\min_{g\in D}\OPT_i(D\setminus\{g\})$ for every agent $i$ and every nonempty target $D$. Between agents, this is the notion called F-EF1 by \citet[Definition~2.9]{dror2021matroid} and FEF1 by \citet[Definition~4.7]{amanatidis2025submodular}. These works do not impose this condition towards the charity: \citet{dror2021matroid} focus on complete allocations, and \citet{amanatidis2026submodular}, in the full version of their work, instead require each agent's utility to be at least a fixed fraction of her value for every feasible subset of the unallocated goods. Both FEF1 and strong FEF1 are implied by the following notion of \citet{barman2023gap}.

\begin{definition}[FEFx]
\label{def:fefx}
For $\gamma\in[0,1]$, an allocation is \emph{$\gamma$-FEFx} if, for every agent $i$, every target $D$, and every $T\subsetneq D$ with $T\in\F_i$, it holds that $u_i\ge\gamma\cdot v_i(T)$. A $1$-FEFx allocation is simply said to be \emph{FEFx}.
\end{definition}

In other words, FEFx requires the envy towards every feasible subset of a target to disappear after the removal of \emph{any} good from the target. \citet{hu2026ef1} refer to this notion as strong FEFx, in order to distinguish it from a weaker variant. We say that an allocation satisfies one of the above notions \emph{between agents} if the respective condition holds for all targets other than the charity. The three exact notions form a hierarchy.

\begin{proposition}
\label{prop:hierarchy}
Every FEFx allocation satisfies strong FEF1, and every allocation satisfying strong FEF1 is FEF1. Neither converse holds, even for one agent and goods of unit size.
\end{proposition}

\begin{proof}
Consider an agent $i$ and a nonempty target $D$. Suppose first that the allocation is FEFx, and let $g\in D$ be arbitrary. Every $T\subseteq D\setminus\{g\}$ is a proper subset of $D$, so $u_i\ge v_i(T)$ whenever $T\in\F_i$; hence $g$ fulfills the requirement of strong FEF1. Suppose next that the allocation satisfies strong FEF1 with good $g\in D$, and consider any nonempty $T\subseteq D$ with $T\in\F_i$. If $g\in T$, then $T\setminus\{g\}\subseteq D\setminus\{g\}$ is feasible for $i$ because $\F_i$ is hereditary, so $u_i\ge v_i(T\setminus\{g\})$. If $g\notin T$, then $u_i\ge v_i(T)\ge v_i(T\setminus\{h\})$ for any $h\in T$. In either case, the FEF1 condition holds for $T$.

For the separations, consider one agent and goods of unit size. First, let the budget be $2$, let there be five goods with values $100,51,51,51,51$, and give the agent the good of value $100$. Every feasible subset of the charity contains at most two goods, so removing its most valuable good leaves a value of at most $51$; hence the allocation is FEF1. However, after the removal of any single good from the charity, two of the remaining goods form a feasible set of value $102>100$, so strong FEF1 fails. Second, let the budget be $1$, let there be three goods with values $1,2,1$, and give the agent the first good. Removing the good of value $2$ from the charity leaves only feasible sets of value at most $1$, so strong FEF1 holds, but the proper subset of the charity consisting of the good of value $2$ violates FEFx.
\end{proof}

Since FEFx allocations always exist under generalized assignment constraints \citep[Theorem~2]{barman2023gap}, Proposition~\ref{prop:hierarchy} implies that allocations satisfying strong FEF1 or FEF1 exist as well. Our question is therefore a computational one. Following \citet{grotschel1988geometric}, an algorithm runs in \emph{strongly polynomial time} if the number of arithmetic operations that it performs is bounded by a polynomial in $n$ and $m$, independently of the numbers in the input, and all numbers arising during its execution have encoding length polynomial in that of the input. Accordingly, our running time bounds count arithmetic operations, and oracle calls are counted separately where stated.

\section{Computing FEF1 Allocations}
\label{sec:algorithm}

In this section, we show that FEF1 allocations can be computed efficiently whenever each agent's value can be approximately maximized over her feasible bundles. Throughout, let $\delta\in(0,1/(2m)]$ be an accuracy parameter, and let $\alpha:=1-\delta$ and $\beta:=\alpha^2$. An \emph{$\alpha$-approximate optimization oracle} for agent $i$ returns, for any $D\subseteq M$, a set $K_i(D)\subseteq D$ such that $K_i(D)\in\F_i$ and $v_i(K_i(D))\ge\alpha\cdot\OPT_i(D)$. We do not require the answers of the oracle to be consistent across different sets $D$. Our main result is the following.

\begin{theorem}
\label{thm:oracle}
Suppose that every agent $i$ has a hereditary family $\F_i$, given by a membership oracle, together with an $\alpha$-approximate optimization oracle. Then an allocation that is FEF1 and $\beta$-FEFx can be computed using at most $J:=n\big(\binom{m}{2}+m+(m-1)\ln 2/\delta\big)$ bundle updates. The algorithm consists of at most $J+1$ iterations, each of which makes at most one update, $2n$ oracle calls, and $2n$ membership queries, and otherwise takes $O(nm)$ time, after a preprocessing step that takes $O(nm\log m)$ time. In particular, for $\delta=1/(2m)$, the algorithm makes $O(nm^2)$ updates and $O(n^2m^2)$ oracle calls.
\end{theorem}

The bounds in Theorem~\ref{thm:oracle} depend only on $n$, $m$, and $\delta$. For comparison, the analysis of \citet[Theorem~4.1]{hu2026ef1} for budget constraints yields $O(nm^3+nm(L+\ln m))$ updates, each preceded by $O(nm^2)$ oracle calls, where $L$ denotes the length of the input. Under generalized assignment constraints, the oracle can be implemented by a profit-scaling scheme for the knapsack problem, which yields the following corollary. We prove both results in Section~\ref{sec:proofs}.

\begin{corollary}
\label{cor:gap}
Under generalized assignment constraints, an FEF1 allocation can be computed in $O(n^2m^5)$ time; in particular, FEF1 allocations can be computed in strongly polynomial time. Moreover, for every $\eta\in(0,1)$, an allocation that is both FEF1 and $(1-\eta)$-FEFx can be computed in $O(n^2(m^5+m^3/\eta^2))$ time.
\end{corollary}

For budget constraints, \citet[Theorem~4.1]{hu2026ef1} bounded the running time of their algorithm by $O(n^2m^9+n^2m^7(L+\ln m))$. The second part of Corollary~\ref{cor:gap} complements the fully polynomial-time approximation scheme of \citet[Theorem~4]{barman2023gap}, which computes a $(1-\eta)$-FEFx allocation. Such an allocation need not be FEF1, because FEFx imposes no condition on a target as a whole. For example, if an agent with budget $2$ receives a good of unit size that she values at $1-\eta$, and the charity consists of two goods of unit size that she values at $1$ each, then the allocation is $(1-\eta)$-FEFx but not FEF1.

\subsection{The Algorithm}

Our algorithm, stated as Algorithm~\ref{alg:fef1}, follows the framework of \citet{hu2026ef1}. It starts with all goods in the charity, and in each iteration, it replaces the bundle of one agent by a subset of the charity and returns her previous bundle to the charity. The utilities $u_i$ are computed at the beginning of each iteration and remain unchanged during it. We say that agent $i$ \emph{desires} a set $T\subseteq M$ if $T\in\F_i$ and $\alpha\cdot v_i(T)>u_i$, i.e., she can accommodate $T$ and values it more than $1/\alpha$ times her current bundle.

To choose the new bundle, the algorithm first looks for an agent $k$ who desires the answer $K_k(C)$ of her oracle on the charity. It then shrinks this \emph{candidate} $T$, in the spirit of the minimal envied subsets of \citet{chaudhury2021charity} and \citet{barman2023gap}, until every agent $i$ values each proper subset of $T$ that is feasible for her at most $1/\beta$ times her utility (Lemma~\ref{lem:candidate}). Our shrinking step rests on two observations. First, if agent $i$ can accommodate $T$, then her most valuable proper subset of $T$ is obtained by removing a good of minimum value to her, since $\F_i$ is hereditary; hence no oracle call is needed. Second, once no feasible proper subset of $T$ is worth much more to agent $i$ than her own bundle, the same holds for every subset of $T$. The algorithm therefore examines each agent only once: if $i$ cannot accommodate $T$, it calls her oracle on $T$ and replaces $T$ by the answer if $i$ desires it, and as long as $i$ desires the set obtained from $T$ by removing a good of minimum value to her, it removes such a good. Whenever the candidate is replaced, the agent $i$ responsible for the replacement becomes the agent $k$ associated with $T$.

The algorithm then examines the agents who can accommodate the entire candidate. If none of them violates the FEF1 condition towards $T$, the candidate is given to $k$, and we call this an \emph{ordinary update}. Otherwise, an agent $r$ violating the condition receives $T$ without a good of minimum value to her, and we call this a \emph{repair} \citep{hu2026ef1}. All ties are broken according to fixed orders on the agents and the goods.

\begin{algorithm}[t]
\caption{Computing an FEF1 allocation}
\label{alg:fef1}
\begin{algorithmic}[1]
\State $A_i\gets\varnothing$ for every $i\in N$, and $C\gets M$
\Loop \label{line:loop}
  \State $u_i\gets v_i(A_i)$ and $Z_i\gets K_i(C)$ for every $i\in N$
  \If{no agent $i$ desires $Z_i$} \label{line:stop}
    \State \Return $(A_1,\dots,A_n)$ \Comment{the charity is $C$}
  \EndIf
  \State $k\gets$ the first agent who desires $Z_k$, and $T\gets Z_k$
  \For{each agent $i\in N$} \label{line:for} \Comment{shrinking}
    \If{$T\notin\F_i$} \label{line:oracle}
      \State $T'\gets K_i(T)$
      \If{$i$ desires $T'$}
        \State $T\gets T'$ and $k\gets i$
      \EndIf
    \EndIf
    \If{$T\in\F_i$}
      \While{$\alpha\cdot\big(v_i(T)-\min_{g\in T}v_i(g)\big)>u_i$} \label{line:remove}
        \State $T\gets T\setminus\{h\}$ for some $h\in\arg\min_{g\in T}v_i(g)$, and $k\gets i$
      \EndWhile
    \EndIf
  \EndFor \label{line:endfor}
  \If{some $r\in N$ satisfies $T\in\F_r$ and $u_r<v_r(T)-\max_{g\in T}v_r(g)$} \label{line:repair}
    \State $S\gets T\setminus\{h_r\}$ for the first such $r$, where $h_r\in\arg\min_{g\in T}v_r(g)$ \Comment{repair}
  \Else
    \State $r\gets k$ and $S\gets T$ \Comment{ordinary update}
  \EndIf
  \State $C\gets(C\setminus S)\cup A_r$ and $A_r\gets S$
\EndLoop
\end{algorithmic}
\end{algorithm}

\subsection{Correctness}

The choice of $\beta$ ensures that a multiplicative guarantee of $\beta$ already implies the FEF1 condition \citep{hu2026ef1}. Indeed, since a maximum is at least an average and $\beta\ge 1-2\delta\ge 1-1/m$, every agent $i$ and every nonempty set $Y\subseteq M$ satisfy
\begin{equation}
\label{eq:slack}
v_i(Y)-\max_{g\in Y}v_i(g) \le \left(1-\frac{1}{|Y|}\right)v_i(Y) \le \left(1-\frac{1}{m}\right)v_i(Y) \le \beta\cdot v_i(Y).
\end{equation}
We begin with the property of the candidate obtained after shrinking. We state it for an arbitrary initial candidate, as we will use it again in Section~\ref{sec:welfare}.

\begin{lemma}
\label{lem:candidate}
Consider an iteration of Algorithm~\ref{alg:fef1} in which the shrinking step starts with a candidate $T\subseteq C$ desired by its associated agent $k$. Then the shrinking step ends with a nonempty candidate $T\subseteq C$ desired by its associated agent $k$, and $u_i\ge\beta\cdot v_i(Y)$ for every agent $i$ and every $Y\subsetneq T$ with $Y\in\F_i$.
Moreover, if the candidate belongs to $\F_i$ at the end of the iteration of the for-loop for agent $i$, then $u_i\ge\alpha\cdot v_i(Y)$ for every $Y\subsetneq T$.
\end{lemma}

\begin{proof}
Every replacement of the candidate preserves the property that its associated agent desires it: an oracle answer obtained in the test of line~\ref{line:oracle} replaces the candidate only if agent $i$ desires it, and a good $h$ is removed in line~\ref{line:remove} only if $T\in\F_i$ and $\alpha\cdot v_i(T\setminus\{h\})>u_i$, so that $i$ desires $T\setminus\{h\}$ because $\F_i$ is hereditary. In both cases, the new candidate is a subset of the previous one. Since $u_k\ge 0$, a desired set is nonempty.

Now fix an agent $i$, and let $U$ be the candidate at the end of the iteration of the for-loop for agent $i$; the final candidate $T$ is a subset of $U$. If $U\in\F_i$, the loop in line~\ref{line:remove} stopped because $\alpha\cdot(v_i(U)-\min_{g\in U}v_i(g))\le u_i$. Every proper subset of $U$ is contained in $U\setminus\{g\}$ for some $g\in U$, whose value is at most $v_i(U)-\min_{g\in U}v_i(g)$, so we obtain $u_i\ge\alpha\cdot v_i(Y)$ for every $Y\subsetneq U$. If $U\notin\F_i$, then the candidate did not change during the iteration for agent $i$, since both kinds of replacement produce a set in $\F_i$. Hence the oracle of agent $i$ was called on $U$ after the test in line~\ref{line:oracle}, and $i$ does not desire its answer $K_i(U)$, and every $Y\subseteq U$ with $Y\in\F_i$ satisfies $u_i\ge\alpha\cdot v_i(K_i(U))\ge\alpha^2\cdot\OPT_i(U)\ge\beta\cdot v_i(Y)$. Since every proper subset of $T$ is a proper subset of $U$ and $\alpha\ge\beta$, the claimed inequalities hold for $T$ in both cases.
\end{proof}

\begin{lemma}
\label{lem:progress}
Each update of Algorithm~\ref{alg:fef1} keeps the allocation feasible, leaves the bundles of all agents other than its recipient $r$ unchanged, and strictly increases $u_r$. More precisely, if $u_r'$ denotes the utility of $r$ after the update and $T$ is the candidate, then $u_r'>u_r/\alpha$ for an ordinary update, and $u_r'-u_r>\max_{g\in T}v_r(g)-\min_{g\in T}v_r(g)\ge 0$ for a repair.
\end{lemma}

\begin{proof}
In an ordinary update, the recipient desires $T$ by Lemma~\ref{lem:candidate}, so $T\in\F_r$ and $u_r'=v_r(T)>u_r/\alpha$. In a repair, the recipient $r$ satisfies $T\in\F_r$, so $S=T\setminus\{h_r\}\in\F_r$ because $\F_r$ is hereditary, and $u_r'-u_r=v_r(T)-\min_{g\in T}v_r(g)-u_r>\max_{g\in T}v_r(g)-\min_{g\in T}v_r(g)$, where the inequality is the repair condition in line~\ref{line:repair}. In both cases, $S\subseteq T\subseteq C$, so returning the previous bundle of $r$ to the charity keeps the bundles disjoint.
\end{proof}

\begin{lemma}
\label{lem:fairness}
If the allocation before an update of Algorithm~\ref{alg:fef1} is FEF1 and $\beta$-FEFx between agents, then so is the allocation after the update. Moreover, if Algorithm~\ref{alg:fef1} returns an allocation that is FEF1 and $\beta$-FEFx between agents, then this allocation is FEF1 and $\beta$-FEFx.
\end{lemma}

\begin{proof}
Consider an update with recipient $r$ and new bundle $S$. Comparisons with all other bundles are unaffected, because these bundles do not change and no utility decreases by Lemma~\ref{lem:progress}. It remains to consider an agent $i\ne r$ and a set $Y\subseteq S$ with $Y\in\F_i$. If $Y\subsetneq T$, then Lemma~\ref{lem:candidate} gives $u_i\ge\beta\cdot v_i(Y)$, which by~\eqref{eq:slack} also implies the FEF1 condition for $Y$ if $Y$ is nonempty. In a repair, $S\subsetneq T$, so this covers every such $Y$. In an ordinary update, the only remaining case is $Y=S=T$, for which $\beta$-FEFx imposes no condition, and since no repair took place, every agent $i$ with $T\in\F_i$ satisfies $u_i\ge v_i(T)-\max_{g\in T}v_i(g)$.

When the algorithm terminates, no agent $i$ desires $K_i(C)$, so $u_i\ge\alpha\cdot v_i(K_i(C))\ge\beta\cdot\OPT_i(C)$ by the test in line~\ref{line:stop}. Hence every $Y\subseteq C$ with $Y\in\F_i$ satisfies $u_i\ge\beta\cdot v_i(Y)$, and~\eqref{eq:slack} yields the FEF1 condition towards the charity.
\end{proof}

Since the initial allocation of Algorithm~\ref{alg:fef1} consists of empty bundles, Lemma~\ref{lem:fairness} implies that the algorithm returns an FEF1 and $\beta$-FEFx allocation whenever it terminates. We remark that the proofs of Lemmas~\ref{lem:candidate}--\ref{lem:fairness} use only that each family $\F_i$ is hereditary and that each oracle satisfies its own approximation guarantee. In particular, they never relate the feasibility of a set for one agent to its feasibility for another agent, which is why the framework of \citet{hu2026ef1} continues to work when sizes differ across agents.

\subsection{Bounding the Number of Updates}

Since each update strictly increases the utility of its recipient and each agent has finitely many possible utilities, Algorithm~\ref{alg:fef1} terminates. However, an agent may have exponentially many possible utilities, so a polynomial bound requires more care. We bound the numbers of repairs and ordinary updates of a fixed agent $i$ separately. Both bounds depend only on the form of the updates---the repair rule and Lemma~\ref{lem:progress}---and not on how the candidates are chosen, so they apply to every finite prefix of an execution. Throughout, let $w_1\le w_2\le\dots\le w_m$ denote the values $v_i(g)$ for $g\in M$ in nondecreasing order, and identify each good with its position in this order.

\citet{hu2026ef1} showed that each repair either moves the utility of $i$ past one of $O(m^2)$ breakpoints, each an integer multiple of some value $w_j$, or increases it by a substantial amount within an interval between two consecutive breakpoints, which yields a bound of $m^3$. We show that the breakpoints alone suffice, provided that they are chosen as sums of consecutive values $w_j+w_{j+1}+\dots+w_{j+\ell-1}$.

\begin{lemma}
\label{lem:windows}
Let $T\subseteq M$ with $|T|=\ell+1\ge 2$. Then there exists $j\in\{2,3,\dots,m-\ell+1\}$ such that
\[
v_i(T)-\max_{g\in T}v_i(g)\ \le\ w_j+w_{j+1}+\dots+w_{j+\ell-1}\ \le\ v_i(T)-\min_{g\in T}v_i(g).
\]
\end{lemma}

For example, let $m=5$ and $(w_1,\dots,w_5)=(1,1,1,4,4)$, and let $T$ consist of the goods at positions $1$, $4$, and $5$, so that $\ell=2$, $v_i(T)-\max_{g\in T}v_i(g)=5$, and $v_i(T)-\min_{g\in T}v_i(g)=8$. The sums of two consecutive values starting at positions $2$, $3$, and $4$ are $2$, $5$, and $8$, so $j=3$ satisfies the lemma. Accordingly, a repair with candidate $T$ that raises the utility of agent $i$ from some $u<5$ to $8$ moves it past $w_3+w_4=5$.

\begin{proof}
The idea is to slide a window of $\ell$ consecutive positions to the right until its sum reaches the left-hand side, and to show that the sum cannot overshoot the right-hand side at that point. Write $T=\{p_0,p_1,\dots,p_\ell\}$ with $p_0<p_1<\dots<p_\ell$, so that the left-hand and right-hand sides of the claim are $a:=\sum_{r=0}^{\ell-1}w_{p_r}$ and $b:=\sum_{r=1}^{\ell}w_{p_r}$. For $j\in[m-\ell+1]$, let $c_j:=\sum_{r=1}^{\ell}w_{j-1+r}$; the sequence $(c_j)$ is nondecreasing. Since $p_r\ge r+1$ for every $r$, we have $c_2=\sum_{r=1}^{\ell}w_{r+1}\le b$, and since $c_{m-\ell+1}$ is the sum of the $\ell$ largest values, $c_{m-\ell+1}\ge a$. Let $j\ge 2$ be the smallest index with $c_j\ge a$. If $j=2$, we are done, so assume that $j\ge 3$ and hence $c_{j-1}<a$.

For $r\in\{0,1,\dots,\ell\}$, let $d_r:=w_{p_r}-w_{j-1+r}$, so that $\sum_{r=0}^{\ell-1}d_r=a-c_{j-1}>0$ and $\sum_{r=1}^{\ell}d_r=b-c_j$. Since $p_{r+1}-p_r\ge 1$, the difference $p_r-(j-1+r)$ is nondecreasing in $r$. Hence, if $d_r>0$ for some $r$, then $p_r>j-1+r$, so $p_s>j-1+s$ and therefore $d_s\ge 0$ for every $s>r$. If $d_0>0$, it follows that $\sum_{r=1}^{\ell}d_r\ge 0$. Otherwise, since $\sum_{r=0}^{\ell-1}d_r>0\ge d_0$, some $d_r$ with $1\le r\le\ell-1$ is positive, so $d_\ell\ge 0$ and $\sum_{r=1}^{\ell}d_r=\sum_{r=0}^{\ell-1}d_r-d_0+d_\ell>0$. In both cases, $c_j\le b$.
\end{proof}

\begin{theorem}
\label{thm:repairs}
Each agent receives at most $m(m-1)/2$ repairs.
\end{theorem}

\begin{proof}
Let $P$ be the set of sums $w_j+w_{j+1}+\dots+w_{j+\ell-1}$ for $1\le\ell\le m-1$ and $2\le j\le m-\ell+1$, so that $|P|\le\sum_{\ell=1}^{m-1}(m-\ell)=m(m-1)/2$. Consider a repair with candidate $T$ that increases the utility of agent $i$ from $u$ to $u'$. By the repair condition, $u<v_i(T)-\max_{g\in T}v_i(g)$, which implies $|T|\ge 2$ since $u\ge 0$, and $u'=v_i(T)-\min_{g\in T}v_i(g)$. By Lemma~\ref{lem:windows}, the interval $(u,u']$ contains an element of $P$. Since the utility of $i$ never decreases, the intervals $(u,u']$ of different repairs are pairwise disjoint, so each element of $P$ lies in at most one of them. Hence there are at most $|P|$ repairs.
\end{proof}

Theorem~\ref{thm:repairs} cannot be improved by more than a constant factor, even for a single agent under budget constraints.

\begin{proposition}
\label{prop:repair-lower}
Let $\delta=1/(2m)$. For infinitely many values of $m$, there exist an instance with one agent under budget constraints and an $\alpha$-approximate optimization oracle for this agent such that Algorithm~\ref{alg:fef1} performs at least $m^2/100$ repairs.
\end{proposition}

We defer the proof to Appendix~\ref{app:lower}. Proposition~\ref{prop:repair-lower} shows that an algorithm with asymptotically fewer repairs would have to exploit the structure of a specific oracle or deviate from the update rule of \citet{hu2026ef1}.

For ordinary updates, \citet{hu2026ef1} combined the multiplicative gain from Lemma~\ref{lem:progress} with the ratio between the agent's total value and her smallest positive value, which yields a bound that depends on the input numbers. We instead group the positive utilities of the agent according to the largest value $w_j$ that does not exceed them. Within each group, the utility can only grow by a bounded factor, and these factors multiply to at most $2^{m-1}$, regardless of how the values compare with each other.

\begin{lemma}
\label{lem:ordinary}
Each agent receives at most $m+(m-1)\ln 2/\delta$ ordinary updates.
\end{lemma}

\begin{proof}
If all values of agent $i$ are zero, she never desires a set and receives no update. Otherwise, let $j_0$ be the smallest index with $w_{j_0}>0$, and for each $j\ge j_0$, let $S_j:=w_1+w_2+\dots+w_j$. For a positive utility $u$ of agent $i$, let $j(u)$ be the largest index $j$ with $w_j\le u$. A bundle of value $u$ contains a good of positive value, and every good in it has value at most $u$ and hence position at most $j(u)$. Therefore, $j(u)\ge j_0$, $w_{j(u)}\le u\le S_{j(u)}$, and $u<w_{j(u)+1}$ if $j(u)<m$.
For $j_0\le j<m$, let $\rho_j:=\min\{w_{j+1},S_j\}/w_j$, and let $\rho_m:=S_m/w_m$; note that $\rho_j\ge 1$ for every $j$.

The index $j(u)$ does not decrease along an execution. At most one ordinary update starts from utility zero, and at most $m-j_0$ ordinary updates between positive utilities change $j(u)$. Now fix $j$, and let $x_1\to y_1,\dots,x_t\to y_t$ be, in chronological order, the ordinary updates of $i$ with $j(x_s)=j(y_s)=j$, where $t\ge 1$. By Lemma~\ref{lem:progress}, we have $y_s/x_s>1/\alpha$ for each $s$. Moreover, $x_{s+1}\ge y_s$ for each $s$, and by the bounds on $u$ in terms of $j(u)$ above, $x_1\ge w_j$ and $y_t\le S_j$, as well as $y_t<w_{j+1}$ if $j<m$. Hence
\[
\alpha^{-t} < \prod_{s=1}^t\frac{y_s}{x_s} \le \frac{y_t}{x_1} \le \rho_j,
\]
so $t<\ln\rho_j/(-\ln\alpha)\le\ln\rho_j/\delta$.

It remains to show that $\prod_{j=j_0}^m\rho_j\le 2^{m-j_0}$. Since $\min\{x,y\}\cdot(x+y)\le 2xy$ for all positive $x$ and $y$, and $S_{j+1}=S_j+w_{j+1}$, we have for $j_0\le j<m$ that
\[
\rho_j=\frac{\min\{w_{j+1},S_j\}}{w_j}\le\frac{2S_jw_{j+1}}{w_jS_{j+1}}=2\cdot\frac{S_j/w_j}{S_{j+1}/w_{j+1}}.
\]
Multiplying these inequalities over $j$, the right-hand sides telescope; since $S_{j_0}=w_{j_0}$ and $\rho_m=S_m/w_m$, we obtain $\prod_{j=j_0}^{m}\rho_j\le 2^{m-j_0}$. Altogether, agent $i$ receives at most $1+(m-j_0)+(m-j_0)\ln 2/\delta\le m+(m-1)\ln 2/\delta$ ordinary updates.
\end{proof}

\subsection{Proofs of Theorem~\ref{thm:oracle} and Corollary~\ref{cor:gap}}
\label{sec:proofs}

\begin{proof}[Proof of Theorem~\ref{thm:oracle}]
By Theorem~\ref{thm:repairs} and Lemma~\ref{lem:ordinary}, every finite prefix of an execution of Algorithm~\ref{alg:fef1} contains at most $J$ updates. Every iteration is finite, since each replacement in the shrinking step removes at least one good from the candidate, and every iteration either performs an update or terminates. Hence Algorithm~\ref{alg:fef1} terminates after at most $J+1$ iterations, and by Lemma~\ref{lem:fairness}, it returns an FEF1 and $\beta$-FEFx allocation. In each iteration, the test in line~\ref{line:stop} makes $n$ oracle calls, the shrinking step makes at most one oracle call and one membership query per agent (after the test in line~\ref{line:oracle}, it is known whether $T\in\F_i$, since oracle answers are feasible), and the test in line~\ref{line:repair} makes at most $n$ membership queries. For the remaining work, we sort the goods once for each agent by value, which takes $O(nm\log m)$ time. Since the candidate only shrinks within an iteration, the goods of minimum value to agent $i$ in her successive candidates can be found by a single scan of her sorted list. Hence the shrinking step, the evaluation of the oracle answers, and the test in line~\ref{line:repair} take $O(nm)$ time per iteration. Finally, for $\delta=1/(2m)$, we have $J=n(m(m-1)/2+m+2m(m-1)\ln 2)=O(nm^2)$.
\end{proof}

For Corollary~\ref{cor:gap}, we implement the oracle by the classical profit-scaling scheme for the knapsack problem \citep{ibarra1975fast}, using a lower bound on the optimum obtained from its fractional relaxation \citep{dantzig1957discrete,lawler1979fast}. We include a proof in Appendix~\ref{app:oracle} to make sure that its running time does not depend on the numbers in the input.

\begin{lemma}
\label{lem:oracle}
Under generalized assignment constraints, an $\alpha$-approximate optimization oracle for any agent can be implemented in $O(m^2/\delta)$ time.
\end{lemma}

\begin{proof}[Proof of Corollary~\ref{cor:gap}]
For the first part, let $\delta=1/(2m)$. By Theorem~\ref{thm:oracle}, there are $O(nm^2)$ iterations. In each iteration, the $2n$ oracle calls take $O(m^2/\delta)=O(m^3)$ time each by Lemma~\ref{lem:oracle}, each membership query takes $O(m)$ time, and the remaining work takes $O(nm)$ time. This gives a total running time of $O(n^2m^5)$. Apart from the numbers in the proof of Lemma~\ref{lem:scaling}, the algorithm only computes sums of input values and multiplies them by $\alpha$, so all numbers have encoding length polynomial in that of the input, and the algorithm runs in strongly polynomial time.

For the second part, let $\delta:=\min\{\eta/2,1/(2m)\}$, so that $\beta\ge 1-2\delta\ge 1-\eta$ and $1/\delta\le 2m+2/\eta$. By Theorem~\ref{thm:oracle}, there are $O(n(m^2+m/\delta))$ iterations, each taking $O(nm^2/\delta)$ time, which gives a total running time of $O(n^2(m^4/\delta+m^3/\delta^2))=O(n^2(m^5+m^4/\eta+m^3/\eta^2))$. Since $m^4/\eta\le(m^5+m^3/\eta^2)/2$, this is $O(n^2(m^5+m^3/\eta^2))$.
\end{proof}

Since the oracle calls dominate the running time, faster approximation schemes for the knapsack problem directly speed up our algorithm. For instance, with the schemes of \citet{chen2024knapsack} and \citet{mao2024knapsack}, which run in $\widetilde{O}(m+1/\delta^2)$ time, the running time for $\delta=1/(2m)$ becomes $\widetilde{O}(n^2m^4)$.

Theorem~\ref{thm:oracle} applies beyond generalized assignment constraints, as long as near-exact optimization over each agent's feasible bundles is possible.

\begin{remark}
\label{rem:matroid}
Suppose that each $\F_i$ is the family of independent sets of a matroid, or more generally the family of common independent sets of two matroids, where the matroids are given by independence oracles and may differ across agents. Since the restriction of a matroid to a subset $D\subseteq M$ is again a matroid, a set of maximum value within $D$ can be computed by the greedy algorithm \citep{edmonds1971matroids} or by weighted matroid intersection \citep{lawler1975matroid,frank1981weighted}, respectively, in strongly polynomial time \citep[see, e.g.,][Chapters~40 and~41]{schrijver2003combinatorial}. Hence, by Theorem~\ref{thm:oracle}, FEF1 allocations can be computed in strongly polynomial time under such constraints as well. We will see in Section~\ref{sec:matroid3} that this is no longer the case for three matroids.
\end{remark}

\section{Limits of Tractability}
\label{sec:hardness}

We now show that the positive results of Section~\ref{sec:algorithm} are close to the limits of tractability. The slack in~\eqref{eq:slack}, which allows Algorithm~\ref{alg:fef1} to work with approximate oracles, comes from removing the most valuable good from a fixed comparison set. Under strong FEF1, the removed good must be chosen before the comparison set, and the quantity $\min_{g\in D}\OPT_i(D\setminus\{g\})$ may be equal to $\OPT_i(D)$. This is the case, for instance, when the target contains several copies of an optimal feasible set---an observation that underlies our reductions.

\subsection{Strong FEF1 Towards the Charity}

\begin{theorem}
\label{thm:hardness}
Unless $\mathrm{P}=\mathrm{NP}$, there is no polynomial-time algorithm that computes an allocation satisfying strong FEF1 under budget constraints. This holds even when there is only one agent and $v_1(g)=s_1(g)$ is a positive integer for every good $g$.
\end{theorem}

Since budget constraints are a special case of generalized assignment constraints, the same conclusion holds for the latter.

\begin{proof}
We reduce from Exact Cover by 3-Sets (X3C), which is NP-complete \citep{karp1972reducibility}, \citep[Problem~SP2]{garey1979computers}. An instance of X3C consists of a positive integer $q$ and a nonempty family $\mathcal{T}$ of $3$-element subsets of $U:=\{0,1,\dots,3q-1\}$, and asks whether some $q$ members of $\mathcal{T}$ partition $U$. Let $b:=q+1$, $H:=b^{3q}$, $R:=\sum_{e=0}^{3q-1}b^e$, and $B:=qH+R$, and for each $t\in\mathcal{T}$, let $w_t:=H+\sum_{e\in t}b^e$. As in the classical reduction from exact cover to the knapsack problem \citep{karp1972reducibility}, the base-$b$ digits of $w_t$ indicate the elements of $t$, while the term $H$ fixes the number of sets. Since $R=(H-1)/(b-1)<H$, we have $B<(q+1)H$, so at most $q$ of the numbers $w_t$ can add up to at most $B$. We first show the following claim, which we will use again in Section~\ref{sec:between}:
\begin{equation}
\label{eq:digits}
\text{for any } t_1,\dots,t_k\in\mathcal{T} \text{ with } k\le q,\quad w_{t_1}+\dots+w_{t_k}=B \text{ if and only if } k=q \text{ and } t_1,\dots,t_k \text{ form an exact cover.}
\end{equation}
Here, $t_1,\dots,t_k$ need not be distinct. If $t_1,\dots,t_q$ form an exact cover, their numbers add up to $qH+R=B$. Conversely, for each $e\in U$, let $c_e$ be the number of indices $s$ with $e\in t_s$, so that $w_{t_1}+\dots+w_{t_k}=kH+\sum_{e\in U}c_eb^e$. Since $c_e\le k\le q<b$ for every $e$, the second term has base-$b$ digits $c_e$, and it is at most $qR=H-1<H$. Comparing with $B=qH+R$, where $R<H$ as well, yields $k=q$ and $\sum_{e\in U}c_eb^e=R$, so $c_e=1$ for every $e\in U$ by the uniqueness of base-$b$ representations. Hence $t_1,\dots,t_q$ form an exact cover.

For each $t\in\mathcal{T}$, we create $q+2$ goods, each of size and value $w_t$. There is one agent, whose budget is $B$. The construction takes polynomial time. Every feasible bundle contains at most $q$ goods, so by~\eqref{eq:digits}, a bundle of size exactly $B$ exists if and only if the X3C instance has an exact cover.

Now suppose that an algorithm returns a bundle $A_1$ for the agent such that the resulting allocation satisfies strong FEF1, and let $C$ be the charity. Since $|A_1|\le q$ while there are at least $q+2$ goods, $C$ is nonempty, so there exists $g\in C$ such that $v_1(A_1)\ge\OPT_1(C\setminus\{g\})$. If the X3C instance has an exact cover $t_1,\dots,t_q$, then $C\setminus\{g\}$ still contains a good created for each $t_s$, because $A_1\cup\{g\}$ contains at most $q+1$ of the $q+2$ goods created for each set. These goods form a feasible bundle of value $B$, so $v_1(A_1)\ge B$; since $v_1(A_1)=s_1(A_1)\le B$, we have $s_1(A_1)=B$. If the X3C instance has no exact cover, no bundle has size $B$. Therefore, checking whether $s_1(A_1)=B$ decides the X3C instance, which is impossible in polynomial time unless $\mathrm{P}=\mathrm{NP}$.
\end{proof}

Two remarks are in order. First, the difficulty is computational rather than existential: for one agent, any feasible bundle of maximum value satisfies strong FEF1, since every feasible subset of the charity is worth at most this maximum. Second, since the numbers in our reduction are exponentially large, Theorem~\ref{thm:hardness} establishes NP-hardness only in the weak sense. This cannot be improved to strong NP-hardness unless $\mathrm{P}=\mathrm{NP}$, because the pseudo-polynomial-time algorithm of \citet[Theorem~3]{barman2023gap} computes an FEFx allocation, which satisfies strong FEF1 by Proposition~\ref{prop:hierarchy}. Theorem~\ref{thm:hardness} also strengthens the result of \citet[Theorem~5]{barman2023gap} that computing an FEFx allocation is NP-hard. Their reduction from the knapsack problem adds a zero-valued good that the agent cannot accommodate, so that FEFx forces her utility to be at least the value of every feasible subset of the rest of the charity. Under strong FEF1, the removed good may be chosen freely---for example, as the most valuable good of the charity---so a different construction is needed; the copies in our reduction ensure that every removal leaves an optimal feasible subset intact.

\subsection{Strong FEF1 Between Agents}
\label{sec:between}

One may wonder whether Theorem~\ref{thm:hardness} relies on the strong comparison with the charity. We show that this is not the case: strong FEF1 remains hard to achieve even if it is only required between agents, while the charity is only compared using FEF1. Some condition towards the charity is needed, since otherwise leaving all goods in the charity would trivially satisfy strong FEF1 between agents.

\begin{theorem}
\label{thm:between}
Unless $\mathrm{P}=\mathrm{NP}$, there is no polynomial-time algorithm that computes an allocation satisfying strong FEF1 between agents and FEF1 towards the charity under generalized assignment constraints. This holds even when every agent values each good at its size.
\end{theorem}

The idea is as follows. We add an agent who can accommodate every bundle and values all goods equally; by FEF1 towards the charity, she must receive roughly half of the goods. The remaining agents implement the reduction of Theorem~\ref{thm:hardness}, each with respect to a different assignment of the goods to the sets of the X3C instance, so that, for at least one of them, the bundle of the added agent still contains goods corresponding to an exact cover after the removal of any good. A polynomial-size family of assignments that works for every large bundle can be obtained from affine maps over a finite field, as in the universal hash functions of \citet{carter1979universal}. For a prime $p$, we write $\mathbb{F}_p$ for the field of integers modulo $p$.

\begin{lemma}
\label{lem:affine}
Let $k$ and $\tau$ be positive integers, let $p\ge 100k\tau^2$ be a prime, and partition $\mathbb{F}_p$ into sets $P_1,\dots,P_\tau$ whose sizes differ by at most one. Then for every $D\subseteq\mathbb{F}_p$ with $|D|\ge p/3$, there exists a map $f(z)=\lambda z+\mu$ with $\lambda,\mu\in\mathbb{F}_p$ and $\lambda\ne 0$ such that $|f(D)\cap P_j|>k$ for every $j\in[\tau]$.
\end{lemma}

\begin{proof}
Choose $\lambda\in\mathbb{F}_p\setminus\{0\}$ and $\mu\in\mathbb{F}_p$ uniformly at random. Since an affine map is determined by the images of two distinct points, for any distinct $y,y'\in\mathbb{F}_p$, the pair $(f^{-1}(y),f^{-1}(y'))$ is uniformly distributed over the ordered pairs of distinct elements of $\mathbb{F}_p$. Hence, with $\rho:=|D|/p\ge 1/3$, each element of $\mathbb{F}_p$ belongs to $f(D)$ with probability $\rho$, and any two distinct elements both belong to $f(D)$ with probability $|D|(|D|-1)/(p(p-1))\le\rho^2$. Fix $j\in[\tau]$, and let $X:=|f(D)\cap P_j|$; note that $|P_j|\ge\lfloor p/\tau\rfloor\ge 100k\tau$. Then $\mathbb{E}[X]=\rho|P_j|\ge|P_j|/3$, and since the events that different elements of $P_j$ belong to $f(D)$ are pairwise nonpositively correlated, $\mathrm{Var}[X]\le\rho(1-\rho)|P_j|\le|P_j|/4$. As $k\le|P_j|/100$, Chebyshev's inequality gives
\[
\Pr[X\le k]\le\Pr\big[|X-\mathbb{E}[X]|\ge|P_j|/3-k\big]\le\frac{|P_j|/4}{(|P_j|/6)^2}=\frac{9}{|P_j|}\le\frac{9}{100k\tau}.
\]
By the union bound over $j\in[\tau]$, the probability that $|f(D)\cap P_j|\le k$ for some $j$ is at most $9/(100k)<1$, so a suitable map exists.
\end{proof}

\begin{proof}[Proof of Theorem~\ref{thm:between}]
We again reduce from X3C, and use $b$, $H$, $R$, $B$, and $w_t$ as in the proof of Theorem~\ref{thm:hardness}. Let $\tau:=|\mathcal{T}|$. By Bertrand's postulate, there is a prime $p$ with $100\tau^2\le p\le 200\tau^2$, which can be found by trial division. Partition $\mathbb{F}_p$ into sets $(P_t)_{t\in\mathcal{T}}$ whose sizes differ by at most one, let $\Lambda$ be the set of all $p(p-1)$ maps $f(z)=\lambda z+\mu$ with $\lambda\ne 0$, and let $r:=4(|\Lambda|q+1)$. The goods are the pairs $(z,\ell)\in\mathbb{F}_p\times[r]$, so that $m=pr$. For each $f\in\Lambda$, there is an agent $f$ with budget $B$, for whom each good $(z,\ell)$ has size and value $w_t$, where $t$ is the set with $f(z)\in P_t$. There is one further agent, $\star$, with budget $m$, for whom every good has size and value $1$. Since $p=O(\tau^2)$, the construction takes polynomial time.

Consider an allocation satisfying strong FEF1 between agents and FEF1 towards the charity. As in the proof of Theorem~\ref{thm:hardness}, every agent $f\in\Lambda$ receives at most $q$ goods. Since every bundle is feasible for agent $\star$, FEF1 towards the charity implies $|A_\star|\ge|C|-1$, and hence $2|A_\star|+1\ge|A_\star|+|C|\ge m-|\Lambda|q$. Let $D\subseteq\mathbb{F}_p$ be the set of elements $z$ such that $A_\star$ contains a good $(z,\ell)$ for some $\ell$. Since there are $r$ goods for each $z$,
\[
|D|\ge\frac{|A_\star|}{r}\ge\frac{m-|\Lambda|q-1}{2r}=\frac{p}{2}-\frac18\ge\frac{p}{3}.
\]
By Lemma~\ref{lem:affine} with $k=1$, some $f\in\Lambda$ satisfies $|f(D)\cap P_t|\ge 2$ for every $t\in\mathcal{T}$. That is, for every $t\in\mathcal{T}$, the bundle $A_\star$ contains two goods of size $w_t$ for agent $f$. Consequently, for every $g\in A_\star$ and every $t\in\mathcal{T}$, the set $A_\star\setminus\{g\}$ contains a good of size $w_t$ for agent $f$.

If the X3C instance has an exact cover $t_1,\dots,t_q$, then for every $g\in A_\star$, the set $A_\star\setminus\{g\}$ contains goods of sizes $w_{t_1},\dots,w_{t_q}$ for agent $f$, which form a feasible bundle of value $B$ for her. Strong FEF1 of agent $f$ towards agent $\star$, whose bundle is nonempty, therefore implies $v_f(A_f)\ge\min_{g\in A_\star}\OPT_f(A_\star\setminus\{g\})\ge B$, and since $v_f(A_f)=s_f(A_f)\le B$, agent $f$ has utility exactly $B$. If the X3C instance has no exact cover, then by~\eqref{eq:digits}, no agent in $\Lambda$ has a bundle of size exactly $B$. Hence checking whether some agent in $\Lambda$ has utility $B$ decides the X3C instance.
\end{proof}

The heterogeneity of the agents' constraints is essential for Theorem~\ref{thm:between}. Indeed, if all agents have the same feasible family, then every bundle $A_j$ is feasible for every agent $i$, so FEF1 towards $A_j$ gives $u_i\ge v_i(A_j)-\max_{g\in A_j}v_i(g)\ge\OPT_i(A_j\setminus\{g^*\})$, where $g^*$ is a good of maximum value to $i$ in $A_j$. Hence FEF1 between agents implies strong FEF1 between agents, and Algorithm~\ref{alg:fef1} computes allocations satisfying the requirements of Theorem~\ref{thm:between}, for example, under budget constraints with equal budgets.

Both hardness results continue to hold if more than one good may be removed.

\begin{corollary}
\label{cor:k-removals}
For every fixed positive integer $k$, Theorems~\ref{thm:hardness} and~\ref{thm:between} continue to hold if, in Definition~\ref{def:strong}, up to $k$ goods may be removed from each target instead of one.
\end{corollary}

\begin{proof}
In the proof of Theorem~\ref{thm:hardness}, create $q+k+1$ goods for each set in $\mathcal{T}$ instead of $q+2$. After at most $q$ goods are allocated to the agent and at most $k$ goods are removed from the charity, at least one good created for each set remains in the charity, and the rest of the proof is unchanged. In the proof of Theorem~\ref{thm:between}, choose a prime $p$ with $100k\tau^2\le p\le 200k\tau^2$ and apply Lemma~\ref{lem:affine} with this $k$. Then $A_\star$ contains $k+1$ goods of size $w_t$ for some agent $f$ and every $t\in\mathcal{T}$, so at least one of them remains after the removal of any $k$ goods.
\end{proof}

\subsection{The Need for Accurate Optimization}
\label{sec:matroid3}

Finally, we show that the accuracy of the optimization required by Theorem~\ref{thm:oracle} is essential, even for FEF1 itself. We begin with intersections of matroids. A \emph{partition matroid} on $M$ is given by a partition of $M$ into parts and a capacity for each part; its independent sets are the sets that contain at most as many goods from each part as its capacity.

\begin{theorem}
\label{thm:matroid3}
Unless $\mathrm{P}=\mathrm{NP}$, there is no polynomial-time algorithm that computes an FEF1 allocation when each agent's feasible bundles are the common independent sets of three partition matroids. This holds even when there is only one agent, she values every good at $1$, and every part of each partition has capacity $1$.
\end{theorem}

\begin{proof}
We reduce from 3-Dimensional Matching (3DM), which is NP-complete \citep{karp1972reducibility}, \citep[Problem~SP1]{garey1979computers}. An instance of 3DM consists of disjoint sets $X$, $Y$, and $Z$ with $|X|=|Y|=|Z|=q$ and a family $E\subseteq X\times Y\times Z$ of triples, and asks whether some $q$ pairwise disjoint triples exist. Let $\nu$ denote the maximum number of pairwise disjoint triples in $E$. For each triple $e\in E$ and each $c\in\{1,2\}$, we create two goods $(e,c,1)$ and $(e,c,2)$. The first partition matroid has one part for each $x\in X$ and $c\in\{1,2\}$, consisting of the goods $(e,c,1)$ and $(e,c,2)$ for all triples $e$ containing $x$; the second and third partition matroids are defined analogously with respect to $Y$ and $Z$. Every part has capacity $1$, and there is one agent, who values every good at $1$.

A bundle is feasible if and only if, for each $c\in\{1,2\}$, its goods of the form $(e,c,\cdot)$ correspond to pairwise disjoint triples, with at most one of the two goods for each triple. Hence every feasible bundle contains at most $\nu$ goods of the form $(e,c,\cdot)$ for each $c$. Let $A_1$ be the bundle of the agent in an FEF1 allocation, and let $C$ be the charity. Since $A_1$ contains at most one of the goods $(e,c,1)$ and $(e,c,2)$ for every $e$ and $c$, the charity contains the other one, so $C$ contains a feasible set of size $2\nu$. FEF1 towards the charity, with unit values, thus implies $|A_1|\ge 2\nu-1$. As $A_1$ contains at most $\nu$ goods of the form $(e,c,\cdot)$ for each $c$, it contains exactly $\nu$ such goods for some $c$, and the corresponding triples form a maximum set of pairwise disjoint triples. Hence the 3DM instance can be decided by checking whether $\nu=q$.
\end{proof}

Since all numbers in the constructed instance are equal to one, Theorem~\ref{thm:matroid3} holds in the strong sense, in contrast to Theorems~\ref{thm:hardness} and~\ref{thm:between}. Using $k+1$ copies $c\in[k+1]$ instead of two shows that the hardness persists if up to $k$ goods may be removed from each comparison set in Definition~\ref{def:fef1}, for any fixed $k$. More generally, the proof shows that, for one agent with unit values, computing an FEF1 allocation is as hard as computing a largest feasible bundle, whenever the class of constraints is closed under taking disjoint copies and duplicating goods. Together with Remark~\ref{rem:matroid}, Theorem~\ref{thm:matroid3} shows that FEF1 allocations can be computed efficiently for intersections of two matroids but not of three, mirroring the complexity of optimization over such constraints.

Theorem~\ref{thm:oracle} requires an oracle with accuracy $1-1/(2m)$, which is available whenever the corresponding optimization problem admits a fully polynomial-time approximation scheme (FPTAS). One may wonder whether a polynomial-time approximation scheme (PTAS), which achieves accuracy $1-\varepsilon$ for every fixed $\varepsilon>0$, also suffices. We show that this is not the case for agents with several budgets---for instance, a branch of a museum may have limited floor space as well as limited staff time for maintaining its exhibits. Under \emph{$d$-dimensional budget constraints}, which extend generalized assignment constraints to $d$ resources, each agent $i$ has a budget vector $B_i\in\mathbb{R}_{\ge 0}^d$ and a size vector $s_i(g)\in\mathbb{R}_{\ge 0}^d$ for each good $g$, and a bundle $S$ is feasible for her if and only if $s_i(S):=\sum_{g\in S}s_i(g)\le B_i$ in every coordinate. For each fixed $d$, maximizing an agent's value over her feasible bundles is the $d$-dimensional knapsack problem, which admits a PTAS \citep{frieze1984knapsack}, but no FPTAS for $d\ge 2$ unless $\mathrm{P}=\mathrm{NP}$ \citep{magazine1984knapsack}.

\begin{theorem}
\label{thm:budgets3}
Unless $\mathrm{P}=\mathrm{NP}$, there is no polynomial-time algorithm that computes an FEF1 allocation under $3$-dimensional budget constraints. This holds even when there is only one agent.
\end{theorem}

\begin{proof}
We again reduce from X3C, and use $b$, $H$, $R$, $B$, and $w_t$ as in the proof of Theorem~\ref{thm:hardness}. Let $J:=B+1$. For each $t\in\mathcal{T}$, we create $q+1$ goods in a set $M_1$, each with size vector $(w_t,0,0)$ and value $w_t$, and $q+1$ goods in a set $M_2$, each with size vector $(0,w_t,J-w_t)$ and value $J$; note that $w_t\le H+R\le B<J$. There is one agent, whose budget vector is $(B,B,qJ-B)$. The construction takes polynomial time.

Consider a feasible bundle $S$, and let $S_1:=S\cap M_1$ and $S_2:=S\cap M_2$. As in the proof of Theorem~\ref{thm:hardness}, $S_1$ contains at most $q$ goods. Moreover, for each good in $M_2$, the last two coordinates of its size vector add up to $J$, while those of the budget vector add up to $qJ$, so $S_2$ contains at most $q$ goods as well. By~\eqref{eq:digits}, the value $v_1(S_1)$, which equals the first coordinate of $s_1(S_1)$, is at most $B$, with equality only if the sets corresponding to the goods in $S_1$ form an exact cover. Similarly, if $|S_2|=q$, then the last two coordinates of $s_1(S_2)$ are at most $B$ and $qJ-B$, respectively, and add up to $qJ$, so the second coordinate equals $B$ and the sets corresponding to the goods in $S_2$ form an exact cover.

If the X3C instance has no exact cover, then $v_1(S_1)\le B-1$ and $|S_2|\le q-1$ for every feasible bundle $S$, so $v_1(S)\le B-1+(q-1)J$. Suppose now that the X3C instance has an exact cover $t_1,\dots,t_q$, and consider any allocation. Since the bundle of the agent contains at most $q$ goods from each of $M_1$ and $M_2$, whereas we created $q+1$ goods in each of these sets for each $t\in\mathcal{T}$, the charity contains, for each $s\in[q]$, a good in $M_1$ and a good in $M_2$ created for $t_s$. These $2q$ goods form a feasible set of value $B+qJ$, whose most valuable good is worth $J$. Hence, if the allocation is FEF1, then $u_1\ge B+(q-1)J$. Therefore, checking whether the utility of the agent is at least $B+(q-1)J$ decides the X3C instance.
\end{proof}

As with Theorem~\ref{thm:hardness}, Theorem~\ref{thm:budgets3} establishes NP-hardness only in the weak sense, and this cannot be improved unless $\mathrm{P}=\mathrm{NP}$: for fixed $d$ and integral sizes and budgets, the $d$-dimensional knapsack problem can be solved by dynamic programming in time polynomial in $m$ and the budgets, so Theorem~\ref{thm:oracle} yields a pseudo-polynomial-time algorithm.

In the oracle model of Theorem~\ref{thm:oracle}, we can determine the required accuracy more precisely. Theorem~\ref{thm:oracle} allows an accuracy of $1-\delta$ for any $\delta\le 1/(2m)$, and we now show that an accuracy of $1-c/m$ does not suffice for any constant $c>2$, so the error of $1/(2m)$ is optimal up to a constant factor. For this result, we count the queries to the membership and optimization oracles, and allow the algorithm to be randomized.

\begin{theorem}
\label{thm:queries}
Let $c>2$ be a constant, and consider instances with one agent and $m$ goods, in which the hereditary family $\F_1$ is accessed through a membership oracle and an $\alpha$-approximate optimization oracle with $\alpha=1-c/m$. For arbitrarily large $m$, every randomized algorithm that computes an FEF1 allocation with probability at least $2/3$ on each such instance makes $2^{\Omega(m)}$ queries on some such instance, where the constant in $\Omega(\cdot)$ depends on $c$.
\end{theorem}

The idea is to add to a fixed family a few feasible bundles, chosen at random from exponentially many possibilities. An oracle with accuracy $1-c/m$ may ignore these bundles, whereas FEF1 forces the agent to receive one of them, so the algorithm has to find them by itself.

\begin{proof}
Let $\xi>0$ be such that $2(1+\xi)<c$. Let $h\ge 2$ be an integer, and let $k:=2h$, $r:=\lceil(1+\xi)h\rceil$, and $m:=4r$. The goods are of $2r$ \emph{types}, with two goods of each type; $r$ of the types are \emph{high} and the other $r$ are \emph{low}. The agent values each good of a high type at $1+1/k$ and each good of a low type at $1$. All bundles in the families defined below contain at most one good of each type, and for such a bundle $S$, we let $H(S)$ and $L(S)$ denote the numbers of its goods of high and low types, respectively. Let $\mathcal{G}$ be the family of bundles $S$ such that either $|S|\le k-2$, $H(S)\le h$, and $L(S)\le h+1$, or $H(S)=h-2$ and $L(S)=h+1$. This family is hereditary, since removing a good from a bundle of the second kind leaves a bundle of size $k-2$ of the first kind. For a set $X$ of $h$ high types and a set $Y$ of $h$ low types, let $\mathcal{H}_{X,Y}$ be the family of bundles all of whose goods have types in $X\cup Y$. In the instance $I_{X,Y}$, the feasible family is $\F_1:=\mathcal{G}\cup\mathcal{H}_{X,Y}$, which is hereditary. To define the optimization oracle, fix a map $K_{\mathcal{G}}$ that assigns to each $D\subseteq M$ a set of maximum value among the sets in $\mathcal{G}$ contained in $D$. On input $D$, the oracle in $I_{X,Y}$ returns $K_{\mathcal{G}}(D)$ if $v_1(K_{\mathcal{G}}(D))\ge\alpha\cdot\OPT_1(D)$, and a feasible subset of $D$ of value $\OPT_1(D)$ otherwise.

We first show that in every FEF1 allocation of $I_{X,Y}$, the bundle of the agent does not belong to $\mathcal{G}$. Since each feasible bundle contains at most one good of each type, the charity contains a good of each type, and hence a set in $\mathcal{H}_{X,Y}$ with one good of each type in $X\cup Y$. This set has value $k+h/k$, and its most valuable good is worth $1+1/k$, so FEF1 towards the charity requires $u_1\ge k-1+(h-1)/k$. On the other hand, each $S\in\mathcal{G}$ satisfies $v_1(S)=|S|+H(S)/k$, which is at most $k-1+(h-2)/k$ if $|S|=k-1$ and at most $k-2+h/k$ otherwise; both bounds are smaller than $k-1+(h-1)/k$.

Next, choose $X$ and $Y$ uniformly at random, and consider the instance $I_{\mathcal{G}}$ in which $\F_1=\mathcal{G}$ and the oracle returns $K_{\mathcal{G}}(D)$ on input $D$; this oracle is exact. We claim that for any fixed query, the probability that it receives different answers in $I_{X,Y}$ and $I_{\mathcal{G}}$ is at most $2/\binom{r}{h}$. Consider first a membership query for a bundle $S$. The answers differ only if $S\in\mathcal{H}_{X,Y}\setminus\mathcal{G}$. Such a bundle satisfies $H(S)\le h$ and $L(S)\le h$, so $S\notin\mathcal{G}$ implies $|S|\ge k-1$, and therefore $H(S)=h$ or $L(S)=h$. In the former case, the high types of the goods in $S$ form the set $X$, and in the latter case, the low types of the goods in $S$ form the set $Y$; each of these events occurs with probability at most $1/\binom{r}{h}$. Consider next an optimization query for a set $D$. The answers differ only if $v_1(K_{\mathcal{G}}(D))<\alpha\cdot\OPT_1(D)$. In this case, a feasible subset $Z$ of $D$ of value $\OPT_1(D)$ belongs to $\mathcal{H}_{X,Y}\setminus\mathcal{G}$, so $|Z|\ge k-1$ as before. If $|Z|=k-1$, then removing any good from $Z$ yields a set in $\mathcal{G}$, so $v_1(K_{\mathcal{G}}(D))\ge v_1(Z)-(1+1/k)$. If $|Z|=k$, then $D$ contains a good of each type in $X\cup Y$. If $D$ also contains a good of a low type outside $Y$, then $D$ contains a set in $\mathcal{G}$ consisting of $h-2$ goods of $Z$ of high types, the $h$ goods of $Z$ of low types, and this good, so $v_1(K_{\mathcal{G}}(D))\ge v_1(Z)-(1+2/k)$. Since $v_1(Z)\ge k-1$, in both cases we have
\[
v_1(K_{\mathcal{G}}(D))\ \ge\ \left(1-\frac{k+2}{k(k-1)}\right)\cdot\OPT_1(D)\ \ge\ \alpha\cdot\OPT_1(D),
\]
where the last inequality holds for all sufficiently large $h$, because $m\le 2(1+\xi)k+4$ and $2(1+\xi)<c$. This contradicts the assumption that the answers differ. Hence, the answers can differ only if $|Z|=k$ and $D$ contains no good of a low type outside $Y$, that is, if the low types of the goods in $D$ form exactly the set $Y$, which occurs with probability at most $1/\binom{r}{h}$.

Finally, suppose that the algorithm makes at most $Q$ queries on every instance, and fix its random choices. Let $A$ be the bundle that it returns for $I_{\mathcal{G}}$, and consider the sequence of queries that it makes for $I_{\mathcal{G}}$, followed by a membership query for $A$. If all of these queries receive the same answers in $I_{X,Y}$ as in $I_{\mathcal{G}}$, then the algorithm behaves identically on $I_{X,Y}$ and returns $A$, where either $A\notin\F_1$ or $A\in\mathcal{G}$; in both cases, it does not return an FEF1 allocation. By the union bound, the algorithm succeeds on $I_{X,Y}$ with probability at most $2(Q+1)/\binom{r}{h}$ over the choice of $X$ and $Y$. Averaging over the random choices of the algorithm, we obtain $X$ and $Y$ such that the algorithm succeeds on $I_{X,Y}$ with probability at most $2(Q+1)/\binom{r}{h}$. Since $\binom{r}{h}\ge(r/h)^h\ge(1+\xi)^h$ and $h=\Omega(m)$, a success probability of $2/3$ requires $Q\ge(1+\xi)^h/3-1=2^{\Omega(m)}$.
\end{proof}

\section{Fairness and Nash Welfare}
\label{sec:welfare}

An FEF1 allocation may be far from efficient. In this section, we show that fairness can be combined with a strong efficiency guarantee. As is common, we measure efficiency by the \emph{Nash welfare} $\NSW(\cA):=(\prod_{i\in N}u_i)^{1/n}$ of an allocation $\cA$, i.e., the geometric mean of the agents' utilities \citep{caragiannis2019unreasonable}.

Our next result converts given bundles into a fair allocation that retains a constant fraction of their Nash welfare. For the main application, each agent $i$ is given a single bundle $O_i$ of an arbitrary allocation $\cO=(O_1,\dots,O_n)$; we allow several bundles per agent, as this will be useful for Theorem~\ref{thm:nsw-opt}. The algorithm runs in two stages. The first stage runs Algorithm~\ref{alg:fef1} with two modifications. First, an iteration starts only if some agent $t$ desires $O\cap C$ for one of her given bundles $O$, in which case the initial candidate is $T:=O\cap C$ with associated agent $t$; we call $t$ the \emph{owner} of the iteration. Second, in an ordinary update, the candidate is given to the owner $t$ if she still desires it, and to its associated agent otherwise. The first stage ends when no agent desires $O\cap C$ for any of her given bundles $O$. The second stage then runs Algorithm~\ref{alg:fef1} from the resulting allocation instead of the empty allocation. The algorithm is stated as Algorithm~\ref{alg:nsw}.

\begin{algorithm}[t]
\caption{Converting given bundles into a fair allocation}
\label{alg:nsw}
\begin{algorithmic}[1]
\State $A_i\gets\varnothing$ for every $i\in N$, and $C\gets M$
\While{some agent desires $O\cap C$ for one of her given bundles $O$} \Comment{first stage}
  \State $u_i\gets v_i(A_i)$ for every $i\in N$
  \State $t\gets$ the first such agent, and $T\gets O\cap C$ for the first such bundle $O$ of $t$ \Comment{owner $t$}
  \State $k\gets t$, and shrink $T$ as in lines~\ref{line:for}--\ref{line:endfor} of Algorithm~\ref{alg:fef1}
  \If{some $r\in N$ satisfies $T\in\F_r$ and $u_r<v_r(T)-\max_{g\in T}v_r(g)$}
    \State $S\gets T\setminus\{h_r\}$ for the first such $r$, where $h_r\in\arg\min_{g\in T}v_r(g)$ \Comment{repair}
  \ElsIf{$t$ desires $T$}
    \State $r\gets t$ and $S\gets T$ \Comment{ordinary update to the owner}
  \Else
    \State $r\gets k$ and $S\gets T$ \Comment{ordinary update}
  \EndIf
  \State $C\gets(C\setminus S)\cup A_r$ and $A_r\gets S$
\EndWhile
\State run Algorithm~\ref{alg:fef1} from line~\ref{line:loop}, starting from the current $(A_1,\dots,A_n)$ and $C$ \Comment{second stage}
\end{algorithmic}
\end{algorithm}

\begin{theorem}
\label{thm:nsw}
Suppose that every agent $i$ has a hereditary family $\F_i$, given by a membership oracle, together with an $\alpha$-approximate optimization oracle. Let $\ell$ be a positive integer, and suppose that each agent $i$ is given bundles $O_i^1,\dots,O_i^\ell\in\F_i$ such that all $n\ell$ given bundles are pairwise disjoint. Then Algorithm~\ref{alg:nsw} computes an allocation $\cA$ that is FEF1 and $\beta$-FEFx and satisfies
\[
\NSW(\cA)\ \ge\ \frac{\alpha}{\ell+1}\left(\prod_{i\in N}\big(v_i(O_i^1)+\dots+v_i(O_i^\ell)\big)\right)^{1/n},
\]
using at most $J$ updates, with at most $2n$ oracle calls per iteration. In particular, for $\ell=1$ and any allocation $\cO$, it holds that $\NSW(\cA)\ge(\alpha/2)\cdot\NSW(\cO)$.
\end{theorem}

\begin{proof}
Every update in the first stage is an update of Algorithm~\ref{alg:fef1} in the sense of Lemmas~\ref{lem:candidate}--\ref{lem:fairness}: the initial candidate is a subset of the charity desired by its associated agent, the shrinking step and the repair test are unchanged, and in an ordinary update, the recipient desires the candidate. Hence both stages preserve FEF1 and $\beta$-FEFx between agents, the final allocation is FEF1 and $\beta$-FEFx, and Theorem~\ref{thm:repairs} and Lemma~\ref{lem:ordinary} bound the number of updates in both stages together by $J$.

It remains to bound the Nash welfare. We claim that, throughout the first stage, every nonempty bundle $A_j$ is contained in a given bundle $O_t^h$ of a unique agent $t$, and
\begin{equation}
\label{eq:owner}
u_t\ge\alpha\cdot v_t(A_j).
\end{equation}
Consider an iteration with owner $t$ and initial candidate $O_t^h\cap C$. Since the candidate only shrinks, the bundle created in this iteration is contained in $O_t^h$, and $t$ is unique because the given bundles are pairwise disjoint. Moreover, every candidate in this iteration belongs to $\F_t$, since $\F_t$ is hereditary, so Lemma~\ref{lem:candidate} gives $u_t\ge\alpha\cdot v_t(Y)$ for every proper subset $Y$ of the final candidate $T$. If the update is a repair, the new bundle is a proper subset of $T$, so~\eqref{eq:owner} holds. If the update is an ordinary update, then either $T$ is given to $t$, in which case~\eqref{eq:owner} is immediate, or $t$ does not desire $T$, i.e., $u_t\ge\alpha\cdot v_t(T)$. Afterwards, \eqref{eq:owner} is maintained, since $u_t$ never decreases and $A_j$ does not change until $j$ receives a new bundle.

Let $\cA'=(A_1',\dots,A_n')$ and $C'$ be the allocation and the charity at the end of the first stage, and for each agent $i$, let $c_i$ be the number of agents $j$ such that $A'_j$ is nonempty and contained in one of the given bundles of $i$. Each good of a given bundle $O_i^h$ is either in $C'$ or in such a bundle $A'_j$, because any nonempty bundle containing this good is contained in some given bundle, which must be $O_i^h$ since the given bundles are pairwise disjoint. Since agent $i$ desires none of the sets $O_i^h\cap C'$, and by~\eqref{eq:owner},
\[
\sum_{h=1}^{\ell}v_i(O_i^h)=\sum_{h=1}^{\ell}v_i(O_i^h\cap C')+\sum_{j}v_i(A'_j)\le\frac{\ell+c_i}{\alpha}\cdot v_i(A'_i),
\]
where the middle sum is over the agents $j$ counted by $c_i$. Since each nonempty bundle is contained in the given bundles of only one agent, $\sum_{i\in N}c_i\le n$, so the inequality of arithmetic and geometric means gives $\prod_{i\in N}(\ell+c_i)\le(\ell+1)^n$. Therefore,
\[
\prod_{i\in N}\sum_{h=1}^{\ell}v_i(O_i^h)\le\left(\frac{\ell+1}{\alpha}\right)^n\prod_{i\in N}v_i(A'_i).
\]
Since the second stage does not decrease any utility, the claimed bound on $\NSW(\cA)$ follows.
\end{proof}

Under generalized assignment constraints, we obtain the following consequence by choosing $\delta$ appropriately and using the oracle of Lemma~\ref{lem:oracle}.

\begin{corollary}
\label{cor:nsw-gap}
Under generalized assignment constraints, for any allocation $\cO$ and any $\varepsilon,\eta\in(0,1)$, an allocation $\cA$ that is FEF1 and $(1-\eta)$-FEFx and satisfies $\NSW(\cA)\ge(1/2-\varepsilon)\cdot\NSW(\cO)$ can be computed in time polynomial in $n$, $m$, $1/\varepsilon$, and $1/\eta$.
\end{corollary}

\begin{proof}
Apply Theorem~\ref{thm:nsw} with $\ell=1$ and $\delta:=\min\{2\varepsilon,\eta/2,1/(2m)\}$, so that $\alpha/2\ge 1/2-\varepsilon$ and $\beta\ge 1-2\delta\ge 1-\eta$. The running time is bounded as in the proof of Corollary~\ref{cor:gap}; testing whether an agent desires $O_i\cap C$ requires no oracle call.
\end{proof}

Our approach adapts that of \citet{barman2024subadditive}, who showed that EF1 allocations retaining half of the optimal Nash welfare exist for subadditive valuations. Since the function $\OPT_i$ is subadditive, and an allocation satisfies strong FEF1 between agents if and only if it is EF1 with respect to the valuations $\OPT_i$, their existence result even yields allocations that satisfy strong FEF1 between agents and retain half of the optimal Nash welfare, after each bundle is replaced by a feasible subset of maximum value. Their polynomial-time algorithm converts any allocation into an EF1 allocation retaining a $1/e^{2/e}\approx 1/2.08$ fraction of its Nash welfare using value queries \citep[Theorem~3.1]{barman2024subadditive}. For the valuations $\OPT_i$, however, each value query requires solving a knapsack problem exactly under generalized assignment constraints. Corollary~\ref{cor:nsw-gap} uses only approximate optimization, retains a fraction approaching $1/2$, and also guarantees FEF1 and $(1-\eta)$-FEFx towards the charity. Between agents, however, it ensures only FEF1, and by Theorem~\ref{thm:between}, strong FEF1 between agents cannot be added to these guarantees in polynomial time unless $\mathrm{P}=\mathrm{NP}$. Moreover, the factor $1/2$ cannot be improved as $\eta$ tends to zero, even in the absence of constraints. Indeed, \citet{feldman2024tradeoffs} showed that for every $\gamma\in(0,1]$ and $\rho>1/(1+\gamma)$, there is an instance with additive valuations in which no allocation, even a partial one, is $\gamma$-EFX and has Nash welfare at least $\rho$ times the maximum. If the budgets are so large that every bundle is feasible, then $\gamma$-FEFx between agents is equivalent to $\gamma$-EFX, so no allocation can be $(1-\eta)$-FEFx and retain a $\rho$ fraction of the maximum Nash welfare for any $\rho>1/(2-\eta)$.

Since Theorem~\ref{thm:nsw} holds for every $\delta\in(0,1/(2m)]$ and there are only finitely many allocations, it also implies that FEFx allocations retaining half of the Nash welfare of any given allocation always exist. This extends the result of \citet{caragiannis2019donating} from the setting without constraints to arbitrary hereditary constraints. If only comparisons between agents are of interest, such allocations can also be obtained by applying the existence result of \citet{barman2024subadditive} for envy-freeness up to any good and subadditive valuations to the valuations $\OPT_i$, as discussed above; Corollary~\ref{cor:fefx-exist} also guarantees FEFx towards the charity.

\begin{corollary}
\label{cor:fefx-exist}
Suppose that every agent $i$ has a hereditary family $\F_i$. Then, for any allocation $\cO$, there exists an FEFx allocation $\cA$ with $\NSW(\cA)\ge\NSW(\cO)/2$.
\end{corollary}

\begin{proof}
For each $\delta\in(0,1/(2m)]$, apply Theorem~\ref{thm:nsw} with $\ell=1$ and exact optimization oracles, which are $\alpha$-approximate for every $\alpha$. This yields a $(1-\delta)^2$-FEFx allocation $\cA^\delta$ with $\NSW(\cA^\delta)\ge((1-\delta)/2)\cdot\NSW(\cO)$. Since there are finitely many allocations, some allocation $\cA$ is equal to $\cA^\delta$ for arbitrarily small $\delta$. Hence, $\cA$ is $\gamma$-FEFx for every $\gamma<1$, i.e., FEFx, and $\NSW(\cA)\ge\NSW(\cO)/2$.
\end{proof}

Theorem~\ref{thm:nsw} requires bundles of high Nash welfare as input. To obtain a guarantee relative to the maximum Nash welfare, we combine it with an approximation algorithm for maximizing the Nash welfare. To this end, we absorb the agents' constraints into their valuations by a fractional relaxation, which allows us to use an algorithm designed for submodular valuations without constraints.

\begin{theorem}
\label{thm:nsw-opt}
Under generalized assignment constraints, for any $\varepsilon,\eta\in(0,1)$, an allocation that is FEF1 and $(1-\eta)$-FEFx and whose Nash welfare is at least $1/(12+\varepsilon)$ times the maximum Nash welfare over all allocations can be computed in time polynomial in the input length, $1/\varepsilon$, and $1/\eta$.
\end{theorem}

\begin{proof}
For each agent $i$ and set $S\subseteq M$, let $S'$ be the set of goods $g\in S$ with $s_i(g)\le B_i$, and let $\phi_i(S)$ be the optimal value of the fractional knapsack problem
\[
\max\Big\{\sum_{g\in S'}v_i(g)x_g\ :\ 0\le x_g\le 1 \text{ for all } g\in S',\ \sum_{g\in S'}s_i(g)x_g\le B_i\Big\}.
\]
The greedy algorithm, which fills the budget with the goods of $S'$ in nonincreasing order of the ratio $v_i(g)/s_i(g)$, solves this problem \citep{dantzig1957discrete}, so $\phi_i$ can be evaluated in polynomial time. Moreover, $\phi_i(X)=v_i(X)$ for every $X\in\F_i$. The function $\phi_i$ is also monotone and submodular. Indeed, the value of any feasible solution $x$ equals $\int_0^\infty\sum_{g\in S':\,v_i(g)/s_i(g)\ge\lambda}s_i(g)x_g\,d\lambda$, and in the greedy solution, the inner sum is the budget used by goods $g$ with $v_i(g)/s_i(g)\ge\lambda$, which equals $\min\{B_i,\,s_i(\{g\in S':v_i(g)/s_i(g)\ge\lambda\})\}$. Hence
\[
\phi_i(S)=\int_0^\infty\min\Big\{B_i,\ s_i\big(\{g\in S':v_i(g)/s_i(g)\ge\lambda\}\big)\Big\}\,d\lambda,
\]
where each integrand is the minimum of a constant and a nonnegative additive function of $S$, and is therefore monotone and submodular.

Let $\tau:=\varepsilon/12$ and $\delta:=\min\{\varepsilon/52,\eta/2,1/(2m)\}$. The algorithm of \citet[Theorem~1.1]{garg2023nsw} computes, in time polynomial in $n$, $m$, and $1/\tau$, disjoint sets $P_1,\dots,P_n\subseteq M$ with $(\prod_{i\in N}\phi_i(P_i))^{1/n}$ at least $1/(4+\tau)$ times the maximum of this quantity over all such sets. Since $\phi_i$ coincides with $v_i$ on feasible bundles, this maximum is at least the maximum Nash welfare $\mathrm{OPT}_{\mathrm{NSW}}$ over all allocations. For each agent $i$, let $O_i^1$ be the set of goods taken entirely by the greedy solution for $\phi_i(P_i)$, and let $O_i^2$ consist of the good taken partially, if any. Both sets belong to $\F_i$, all $2n$ sets are pairwise disjoint, and $\phi_i(P_i)\le v_i(O_i^1)+v_i(O_i^2)$. Applying Theorem~\ref{thm:nsw} with $\ell=2$ and the oracle of Lemma~\ref{lem:oracle}, we obtain an FEF1 and $\beta$-FEFx allocation $\cA$ with
\[
\NSW(\cA)\ \ge\ \frac{\alpha}{3}\left(\prod_{i\in N}\phi_i(P_i)\right)^{1/n}\ \ge\ \frac{1-\delta}{3(4+\tau)}\cdot\mathrm{OPT}_{\mathrm{NSW}}.
\]
Since $(12+\varepsilon)(1-\delta)\ge 12+\varepsilon-13\delta\ge 12+3\varepsilon/4\ge 3(4+\tau)$, this is at least $\mathrm{OPT}_{\mathrm{NSW}}/(12+\varepsilon)$. Moreover, $\beta\ge 1-2\delta\ge 1-\eta$. Finally, since $1/\delta\le 52/\varepsilon+2/\eta+2m$, the algorithm of Theorem~\ref{thm:nsw} runs in time polynomial in the input length, $1/\varepsilon$, and $1/\eta$, as in the proof of Corollary~\ref{cor:gap}.
\end{proof}

The proof of Theorem~\ref{thm:nsw-opt} uses the algorithm of \citet{garg2023nsw} only through evaluations of the functions $\phi_i$. In the same way, any polynomial-time $r$-approximation algorithm for maximizing the Nash welfare under monotone submodular valuations given by value oracles yields an allocation that is FEF1 and $(1-\eta)$-FEFx and retains a $(1-\delta)/(3r)$ fraction of the maximum Nash welfare. \citet{bei2026nsw} recently improved the approximation ratio for submodular valuations to $3.56+\varepsilon$. Their algorithm is randomized and runs in time $n^{\mathrm{poly}(1/\varepsilon)}$, so we state Theorem~\ref{thm:nsw-opt} with the deterministic algorithm of \citet{garg2023nsw}, whose running time is polynomial in $1/\varepsilon$.

We conclude this section by showing that the order of quantifiers, which separates FEF1 from strong FEF1 computationally, also matters for welfare. By Corollary~\ref{cor:fefx-exist} and Proposition~\ref{prop:hierarchy}, there always exists an allocation that satisfies strong FEF1 and retains half of the maximum Nash welfare. Our next result shows that this cannot be improved, even in instances where some FEF1 allocation maximizes the Nash welfare. Recall from Section~\ref{sec:between} that if all agents have the same feasible bundles---for instance, under budget constraints with equal budgets---then FEF1 between agents implies strong FEF1 between agents. By contrast, since the goods are indivisible, a small difference between the budgets can determine whether an agent can accommodate a particular pair of goods, and this suffices for strong FEF1 to lose half of the Nash welfare.

\begin{theorem}
\label{thm:nsw-strong}
For every $\varepsilon>0$ and $\lambda>1$, there exists an instance with budget constraints, in which the largest budget is less than $\lambda$ times the smallest budget, such that some FEF1 allocation maximizes the Nash welfare, but every allocation satisfying strong FEF1 between agents has Nash welfare at most $1/2+\varepsilon$ times the maximum.
\end{theorem}

In our construction, each agent $i\ge 2$ obtains her maximum utility from a pair of goods, while agent $i-1$ values each good of this pair but can accommodate only one of them. Under FEF1, agent $i-1$ compares her bundle only with single goods of the pair, so agent $i$ may receive the pair. Under strong FEF1, however, agent $i-1$ must not envy the good that remains after the removal of either good from the pair. This forces agent $i-1$ to receive her own pair as well, and the requirement propagates along the chain of agents until it reaches an agent without a pair.

\begin{proof}
Let $n\ge 2$, $a\in(1/\sqrt{2},1)$, and $\zeta\in(0,1/(2n))$. The goods are partitioned into sets $G_1,\dots,G_n$, where $G_1$ consists of one good and each of $G_2,\dots,G_n$ consists of two goods. Every good in $G_i$ has size $1+i\zeta$, and agent $i$ has budget $B_i:=2(1+i\zeta)$, so the largest budget is less than $1+n\zeta$ times the smallest one. Agent $i$ values each good in $G_i$ at $a$, each good in $G_{i+1}$ at $1$ if $i<n$, and all other goods at $0$.

Since $3>2(1+n\zeta)$, every feasible bundle contains at most two goods. Agent $i$ can accommodate $G_i$, but since $(1+(i+1)\zeta)+(1+i\zeta)>B_i$, she cannot accommodate a good in $G_{i+1}$ together with any other good in $G_i\cup G_{i+1}$. Hence, the value of a feasible bundle to agent $i$ is $1$ if the bundle contains a good in $G_{i+1}$, and $a$ times the number of its goods in $G_i$ otherwise.

Consider the allocation $\cO$ with $O_i=G_i$ for every agent $i$. For every agent $i$ and every $j\ne i$, each subset of $O_j$ that agent $i$ can accommodate contains at most one good that she values positively, and the charity is empty, so $\cO$ is FEF1. Moreover, $\cO$ maximizes the Nash welfare. Indeed, in any allocation, every agent $i\ge 2$ has utility at most $2a$. If agent $1$ has utility greater than $a$, then she receives a good in $G_2$, so agent $2$ does not receive both goods in $G_2$ and has utility at most $1$; the product of the utilities is then at most $(2a)^{n-2}<a(2a)^{n-1}$, where we use $2a^2>1$. Otherwise, the product is at most $a(2a)^{n-1}$, which is the product for $\cO$.

Now consider an allocation $\cA$ satisfying strong FEF1 between agents. We claim that $A_i\ne G_i$ for every $i\ge 2$. Suppose otherwise, and let $i\ge 2$ be the smallest index with $A_i=G_i$. After the removal of either good from $A_i$, the remaining good can be accommodated by agent $i-1$ and is worth $1$ to her, so strong FEF1 implies $u_{i-1}\ge 1$. Since both goods in $G_i$ belong to $A_i$ and $a<1$, this requires $A_{i-1}=G_{i-1}$, which is impossible if $i-1=1$ because $G_1$ consists of one good, and contradicts the choice of $i$ otherwise. Hence, every agent $i\ge 2$ receives at most one good in $G_i$, so $u_i\le 1$ for every $i<n$ and $u_n\le a$, and therefore $\NSW(\cA)\le a^{1/n}$. Since $\NSW(\cO)=(a(2a)^{n-1})^{1/n}$, we obtain $\NSW(\cA)/\NSW(\cO)\le(2a)^{-(n-1)/n}\le 2^{1/n}/(2a)$. Choosing $n$ sufficiently large, $a$ sufficiently close to $1$, and $\zeta\le(\lambda-1)/n$ completes the proof.
\end{proof}

Our construction adapts an instance of \citet[Theorem~4.3]{barman2024subadditive}, who showed that EF1 allocations may retain only about half of the maximum Nash welfare under subadditive valuations. Since strong FEF1 between agents amounts to EF1 with respect to the valuations $\OPT_i$, Theorem~\ref{thm:nsw-strong} shows that this loss already occurs for the subadditive valuations $\OPT_i$ arising from budget constraints, while FEF1 entails no loss of welfare in the same instances.

\section{Conclusion}
\label{sec:conclusion}

In this paper, we have studied the computation of FEF1 allocations under generalized assignment constraints, which capture settings where the size of a good depends on its recipient. We showed that a variant of the algorithm of \citet{hu2026ef1} computes FEF1 allocations under arbitrary hereditary constraints equipped with sufficiently accurate optimization oracles, and that each agent's bundle is updated $O(m^2)$ times, independently of the numbers in the input; under generalized assignment constraints, this yields a strongly polynomial algorithm with running time $O(n^2m^5)$. The computed allocations can also be made approximately FEFx and retain nearly half of the Nash welfare of any given allocation. On the negative side, strong FEF1 is NP-hard to compute, both for one agent and when it is only required between agents, and may cost half of the Nash welfare even when FEF1 costs nothing. Moreover, FEF1 itself is NP-hard for intersections of three matroids as well as for three budgets per agent, and an oracle accuracy of $1-c/m$ with $c>2$ does not suffice. Overall, our results indicate that the tractability of FEF1 rests on two features: the order of quantifiers in its definition, which is immaterial in the absence of constraints, and the possibility of near-exact optimization over each agent's feasible bundles.

Our work leaves several open questions for future study. A technical question is whether FEF1 allocations can be computed under generalized assignment constraints with asymptotically fewer updates than Algorithm~\ref{alg:fef1}. By Proposition~\ref{prop:repair-lower}, this algorithm may perform $\Omega(m^2)$ repairs even for one agent, so such an improvement would require exploiting the structure of the knapsack problem or deviating from the update rule of \citet{hu2026ef1}. Besides this, an intriguing direction is to determine the complexity of strong FEF1 between agents under budget constraints with different budgets: Theorem~\ref{thm:between} relies on agent-specific sizes, whereas equal budgets make the problem tractable. Another avenue is to characterize the hereditary constraints under which FEF1 allocations can be computed efficiently. Theorem~\ref{thm:oracle} requires an oracle with accuracy $1-1/(2m)$, which is available whenever exact or fully polynomial-time approximate optimization is possible, and Theorems~\ref{thm:budgets3} and~\ref{thm:queries} show that a polynomial-time approximation scheme does not suffice in general. For instance, the complexity of computing FEF1 allocations is unknown when each agent has two budgets, in which case optimization admits a polynomial-time approximation scheme but no fully polynomial-time approximation scheme unless $\mathrm{P}=\mathrm{NP}$ \citep{magazine1984knapsack}. Moreover, Theorems~\ref{thm:oracle} and~\ref{thm:queries} leave a constant-factor gap in the required error, between $1/(2m)$ and $2/m$. The question of whether FEFx allocations can be computed in polynomial time under matroid constraints, posed by \citet{barman2023gap}, also remains open, even though Theorem~\ref{thm:oracle} yields FEF1 together with approximate FEFx in this setting. Finally, it would be interesting to improve the factor $1/(12+\varepsilon)$ in Theorem~\ref{thm:nsw-opt}; by Theorem~\ref{thm:nsw}, a factor close to $1/2$ would follow from a near-optimal approximation of the maximum Nash welfare under generalized assignment constraints. Progress on these questions would help bring the theory of constrained fair division closer to practice.

\bibliographystyle{plainnat}
\bibliography{arxiv}
\clearpage
\appendix

\section{The Profit-Scaling Oracle}
\label{app:oracle}

In this appendix, we verify that the classical profit-scaling scheme for the knapsack problem \citep{ibarra1975fast,lawler1979fast} runs in time that does not depend on the numbers in the input. Lemma~\ref{lem:oracle} is an immediate consequence of the following lemma.

\begin{lemma}
\label{lem:scaling}
Let $\delta\in(0,1)$, and consider generalized assignment constraints. For any agent $i$ and set $D\subseteq M$, a set $K\subseteq D$ with $K\in\F_i$ and $v_i(K)\ge(1-\delta)\cdot\OPT_i(D)$ can be computed in $O(m^2/\delta)$ time.
\end{lemma}

\begin{proof}
Let $D'\subseteq D$ consist of the goods $g\in D$ with $v_i(g)>0$ and $s_i(g)\le B_i$. No good with $s_i(g)>B_i$ belongs to a feasible set, and removing zero-valued goods from a feasible set keeps it feasible without changing its value, so $\OPT_i(D')=\OPT_i(D)$. If $D'=\varnothing$, we return $\varnothing$. Otherwise, order $D'$ by nonincreasing ratio $v_i(g)/s_i(g)$, let $G$ be the longest prefix of this order with $s_i(G)\le B_i$, and let $V:=\max\{v_i(G),\max_{g\in D'}v_i(g)\}$. Since $G$ and every single good of $D'$ are feasible, $V\le\OPT_i(D)$. Moreover, $\OPT_i(D)$ is at most the optimal value of the fractional relaxation, which is at most $v_i(G)$ plus the value of the next good in the order (if any), so $\OPT_i(D)\le 2V$.

Let $\Delta:=\delta V/m$, and for each $g\in D'$, let $p_g:=\lfloor v_i(g)/\Delta\rfloor$, which is an integer between $0$ and $m/\delta$ because $v_i(g)\le V$. Every feasible $X\subseteq D'$ satisfies $\sum_{g\in X}p_g\le v_i(X)/\Delta\le 2m/\delta$, so it suffices to consider rounded profits up to $\Pi:=\lfloor 2m/\delta\rfloor$. Write $D'=\{g_1,\dots,g_{m'}\}$, and for each $j\in\{0,1,\dots,m'\}$ and each integer $\pi\in[0,\Pi]$, let $\sigma(j,\pi)$ be the minimum of $s_i(X)$ over all sets $X\subseteq\{g_1,\dots,g_j\}$ with $\sum_{g\in X}p_g=\pi$, or $\infty$ if there is no such set. Then $\sigma(0,0)=0$, $\sigma(0,\pi)=\infty$ for $\pi>0$, and
\[
\sigma(j,\pi)=\min\bigl\{\sigma(j-1,\pi),\ s_i(g_j)+\sigma(j-1,\pi-p_{g_j})\bigr\},
\]
where the second term is omitted if $\pi<p_{g_j}$. The table has $O(m^2/\delta)$ entries, each computed in constant time, and storing the minimizing choices allows us to reconstruct a set attaining each entry. Let $\pi^*$ be the largest $\pi$ with $\sigma(m',\pi)\le B_i$, and let $K$ be a corresponding set; then $K\in\F_i$.

Let $X^*\subseteq D'$ be a feasible set with $v_i(X^*)=\OPT_i(D)$. Since the rounded profit of $X^*$ is at most $\Pi$ and $s_i(X^*)\le B_i$, the choice of $\pi^*$ gives $\sum_{g\in K}p_g\ge\sum_{g\in X^*}p_g$. Since $\Delta p_g\le v_i(g)<\Delta(p_g+1)$ for every $g\in D'$ and $|X^*|\le m$,
\[
v_i(K)\ge\Delta\sum_{g\in K}p_g\ge\Delta\sum_{g\in X^*}p_g\ge\sum_{g\in X^*}\left(v_i(g)-\Delta\right)\ge\OPT_i(D)-m\Delta=\OPT_i(D)-\delta V\ge(1-\delta)\cdot\OPT_i(D).
\]
Finally, sorting takes $O(m\log m)$ time, which is dominated by the $O(m^2/\delta)$ time for the table. All numbers in this computation are sums of input values or sizes, the quantities $V$ and $\Delta$, or integers between $0$ and $2m/\delta$, so they have encoding length polynomial in that of the input and of $\delta$.
\end{proof}

\section{Proof of Proposition~\ref{prop:repair-lower}}
\label{app:lower}

Let $\kappa$ be a positive integer, $m:=10\kappa$, $\delta:=1/(2m)$, and $\zeta:=1/(100m^3)$. There is one agent, every good has size $1$, and the budget is $B_1:=2\kappa+1$. The goods consist of $2\kappa$ \emph{padding} goods and $4\kappa$ \emph{low} goods, each of value $1$, and $4\kappa$ \emph{high} goods, each of value $1+\zeta$. The fixed order on the goods lists the padding goods first; whenever we select goods of some kind below, we take the first available ones in this order.

We define the oracle $K_1$ as follows. If $|D|\le B_1$, let $K_1(D):=D$, which is optimal since $D$ is feasible. Suppose now that $|D|>B_1$ and that $D$ contains all padding goods, and let $k$ and $t$ be the numbers of all goods and of high goods outside $D$, respectively. We define a set $T\subseteq D$ in three cases, provided that $D$ contains the required goods:
\begin{itemize}
\item if $k=0$, then $T$ consists of one low good;
\item if $1\le k\le 2\kappa$, $t$ is even, and $t+2\le k$, then $T$ consists of $t+2$ high goods and $k-t-1$ low goods;
\item if $1\le k<2\kappa$ and $t=2\lfloor k/2\rfloor$, then $T$ consists of $k+1$ low goods.
\end{itemize}
These cases are disjoint, since $t=2\lfloor k/2\rfloor$ implies $t+2>k$. In each case, $|T|\le k+1\le B_1$, and we let $K_1(D)$ consist of $T$ together with the first $B_1-|T|$ padding goods. In all other cases, $K_1(D)$ consists of the first $B_1$ goods of $D$. This oracle is $\alpha$-approximate: if $|D|>B_1$, then $K_1(D)$ consists of $B_1$ goods of value at least $1$ each, whereas $\alpha\cdot\OPT_1(D)\le(1-\delta)(1+\zeta)B_1<B_1$ since $\zeta<\delta$.

We follow the execution of Algorithm~\ref{alg:fef1} until the agent receives $2\kappa$ high goods, and show by induction that, until then, her bundle always consists of $t$ high goods and $k-t$ low goods for some $k\in\{0,1,\dots,2\kappa\}$ and some even $t$. Such a bundle contains no padding goods, so the oracle is called on a charity $C$ that contains all padding goods and satisfies $|C|\ge m-B_1>B_1$, and the numbers $k$ and $t$ in the definition of $K_1(C)$ are the numbers of all goods and of high goods in the agent's bundle. In the first iteration, $k=0$, and the agent desires $K_1(C)$, which is feasible. In the shrinking step, the candidate is feasible, so no oracle call is made, and the goods of minimum value are the padding goods and the low goods. Since padding goods come first in the fixed order, they are removed one by one, as each removal leaves a superset of $T$, which the agent desires. Once the candidate equals $T$, removing a good of minimum value leaves the empty set, so the shrinking stops. The repair condition fails since $|T|=1$, and an ordinary update gives the agent one low good.

Now suppose that the agent's bundle consists of $t$ high goods and $k-t$ low goods, where $1\le k\le 2\kappa$, $t$ is even, and $t<2\kappa$, so that her utility is $u=k+t\zeta$. We distinguish two cases; in each of them, the new bundle again has the stated form.
\begin{itemize}
\item If $t+2\le k$, the charity contains at least $4\kappa-t\ge t+2$ high goods and $4\kappa-(k-t)\ge k-t-1$ low goods, so $T$ consists of $t+2$ high goods and $k-t-1\ge 1$ low goods, and $v_1(T)=k+1+(t+2)\zeta$. As before, the shrinking step removes all padding goods, since $\alpha\cdot v_1(T)-u=1+2\zeta-\delta(k+1)-\delta(t+2)\zeta>0$, where we use $\delta(k+1)\le(2\kappa+1)/(20\kappa)<1/2$ and $\delta(t+2)\zeta<\zeta$. Once the candidate equals $T$, removing a low good would leave the value $k+(t+2)\zeta$, and $\alpha(k+(t+2)\zeta)-u\le 2\zeta-\delta k<0$ since $2\zeta<\delta$, so the shrinking stops. The repair condition holds, as $u=k+t\zeta<v_1(T)-(1+\zeta)=k+(t+1)\zeta$, so the agent receives $T$ without a low good, i.e., $t+2$ high goods and $k-t-2$ low goods.
\item If $t+2>k$, i.e., $t=2\lfloor k/2\rfloor$, and $k<2\kappa$, then $T$ consists of $k+1$ low goods, which are available since the agent holds at most one low good. The shrinking step again removes all padding goods, since $\alpha(k+1)-u=1-\delta(k+1)-t\zeta>0$. Once the candidate equals $T$, removing a good leaves the value $k\le u$, so the shrinking stops. The repair condition fails, as $v_1(T)-1=k\le u$, so an ordinary update gives the agent $k+1$ low goods.
\end{itemize}
Hence, for each $k\in[2\kappa]$, after the agent receives $k$ low goods, she receives $\lfloor k/2\rfloor$ repairs before receiving $k+1$ low goods, or, for $k=2\kappa$, before receiving $2\kappa$ high goods, at which point the part of the execution that we analyze ends. The total number of repairs is therefore at least $\sum_{k=1}^{2\kappa}\lfloor k/2\rfloor=\kappa^2=m^2/100$.

\end{document}